\documentclass[a4paper,UKenglish,cleveref, autoref, thm-restate]{lipics-v2021}

\nolinenumbers

\usepackage[utf8]{inputenc}
\usepackage[T1]{fontenc}
\usepackage{lmodern}

\usepackage{amsmath} 
\usepackage{amssymb}
\usepackage{mathrsfs}
\usepackage{amsthm}
\usepackage{amsfonts}
\usepackage{graphicx}
\usepackage{algorithm}
\usepackage{algorithmic}
\usepackage{hyperref}
\usepackage{stmaryrd}
\usepackage{mathtools}
\usepackage{todonotes}
\hypersetup{hidelinks,breaklinks}
\usepackage{quiver}

\usepackage[notion, quotation, composition]{knowledge}

\newif\ificalp
\icalpfalse % or \icalpfalse

\knowledge{notion}
|empty word

\knowledge{notion}
|tree
|trees
|$\Sigma$-tree
|$I$-tree
|$[1,2j]$-tree

\knowledge{notion}
|arity
|arities

\knowledge{notion}
|word
|words

\knowledge{notion}
|alphabet
|alphabets

\knowledge{notion}
|concatenation

\knowledge{notion}
|product
|tree product

\knowledge{notion}
|branch
|branches

\knowledge{notion}
|path@tree-p
|paths@tree-p

\knowledge{notion}
|path@automaton-p
|paths@automaton-p

\knowledge{notion}
|index

\knowledge{notion}
|parity accepting
|parity rejecting
|accepting
|rejecting
|acceptance
|accepted

\knowledge{notion}
|accepted@tree

\knowledge{notion}
|acceptance game

\knowledge{notion}
|priority
|priorities

\knowledge{notion}
|parity tree automaton
|tree automata
|automaton
|automata
|non-deterministic $I$-parity tree automaton
|$I$-automaton
|$J$-automaton
|$[1,2j^*]$-automaton
|non-deterministic automaton
|nondeterministic parity tree automaton
|nondeterministic $I$-parity tree automaton

\knowledge{notion}
|state
|states

\knowledge{notion}
|transition
|transitions

\knowledge{notion}
|complete

\knowledge{notion}
|transition path

\knowledge{notion}
|reachable
|reachability
|non-reachable

\knowledge{notion}
|acceptation game
|acceptation games

\knowledge{notion}
|recognized
|recognizes
|recognizing

\knowledge{notion}
|run
|runs

\knowledge{notion}
|accepting run
|accepting@run-acc
|accepting runs
|rejecting run

\knowledge{notion}
|language
|languages

\knowledge{notion}
|$\omega$-regular tree language
|$\omega$-regular tree languages
|regular tree language
|regular language

\knowledge{notion}
|$I$-feasible
|$J$-feasible
|feasible
|feasibility
|$I$-feasibility
|$J$-feasibility

\knowledge{notion}
|deterministic
|deterministic automaton

\knowledge{notion}
|guidable automaton
|guidable automata
|Guidable automata
|guidable
|guidability
|Guidability

\knowledge{notion}
|guiding function

\knowledge{notion}
|preserve acceptance
|preserves acceptance

\knowledge{notion}
|guides
|guided by
|guiding
|guided

\knowledge{notion}
|parity graph
|parity graphs
|$I$-graph
|$I$-graphs

\knowledge{notion}
|even
|odd
|even graph
|odd graph

\knowledge{notion}
|game
|games
|parity game
|parity games

\knowledge{notion}
|play
|plays

\knowledge{notion}
|winning
|winning play
|losing
|losing play

\knowledge{notion}
|strategy
|strategies

\knowledge{notion}
|consistent with

\knowledge{notion}
|winning strategy

\knowledge{notion}
|determined
|determinacy

\knowledge{notion}
|attractor
|attractors

\knowledge{notion}
|attractor decomposition
|attractor decompositions
|decomposition
|decompositions
|($h$-height, $\kappa$-width)-attractor decomposition

\knowledge{notion}
|tight
|tightness

\knowledge{notion}
|well-indexed

\knowledge{notion}
|tree-shape
|tree-shapes

\knowledge{notion}
|level

\knowledge{notion}
|ordered tree
|ordered trees

\knowledge{notion}
|finite tree
|finite
|finite trees

\knowledge{notion}
|child
|children

\knowledge{notion}
|node
|nodes

\knowledge{notion}
|root

\knowledge{notion}
|depth

\knowledge{notion}
|distance
|distances

\knowledge{notion}
|height
|heights

\knowledge{notion}
|diameter
|diameters

\knowledge{notion}
|subtree
|subtrees

\knowledge{notion}
|sibling
|siblings

\knowledge{notion}
|leaf
|leaves

\knowledge{notion}
|mapping
|mappings
|mapping of $G$ into $T$
|mapping of $G$ into $T^*$

\knowledge{notion}
|offset
|offsets

\knowledge{notion}
|$n$-Strahler number
|Strahler number
|Strahler numbers

\knowledge{notion}
|$n$-Strahler number@game-Strahler

\knowledge{notion}
|universal
|universal for $\T$
|universality

\knowledge{notion}
|universal tree
|universal trees

\knowledge{notion}
|width

\knowledge{notion}
|isomorphically embedded
|isomorphic embedding

\knowledge{notion}
|smallest common ancestor
|smallest common ancestors

\knowledge{notion}
|$J,n$-priority transduction game
|priority transduction games
|priority transduction game
|transduction game
|transduction games
|counter
|counters
|register
|registers
|Registers

\knowledge{notion}
|$n$-bound
|$n$-bound by
|$n$-boundedness

\knowledge{notion}
|finitely branching
|finitely-branching

\knowledge{notion}
|rank

\knowledge{notion}
|star-rank

\knowledge{notion}
|regular
|regular tree
|regular trees

\knowledge{notion}
|universal for an automaton $\A$
|universal for $A$
|universal@automaton-univ

\newcommand{\NN}{\mathbb{N}}

\newcommand{\A}{{A}}

\newcommand{\C}{\mathcal{C}}
\newcommand{\D}{\mathcal{D}}

\newcommand{\F}{\mathcal{F}}
\newcommand{\G}{\mathcal{G}}

\renewcommand{\L}{\mathcal{L}}

\newcommand{\R}{\mathcal{R}}
\renewcommand{\S}{\mathcal{S}}
\newcommand{\T}{\mathcal{T}}
\newcommand{\U}{\mathcal{U}}
\newcommand{\V}{\mathcal{V}}

\renewcommand{\leq}{\leqslant}
\renewcommand{\geq}{\geqslant}

\newcommand{\ptime}{{\sc PTime}}

\knowledgenewcommand{\concat}{\cmdkl{\cdot}}
\knowledgenewcommand{\Tr}[1]{\cmdkl{Tr_{#1}}}
\knowledgenewcommand{\trProd}{\cmdkl{\otimes}}
\knowledgenewcommand{\Lang}{\cmdkl{\L}}

\knowledgenewcommand{\attr}{\cmdkl{\mathtt{Attr}}}
\knowledgenewcommand{\restrict}{\cmdkl{\upharpoonright}}

\knowledgenewcommand{\subt}{\cmdkl{\sqsubseteq}}
\knowledgenewcommand{\h}{\cmdkl{\mathtt{h}}}
\knowledgenewcommand{\dist}{\cmdkl{\delta}}

\knowledgenewcommand{\Ut}[1]{\cmdkl{\U_{#1}}}

\knowledgenewcommand{\Reg}[2]{\cmdkl{\T_{#1}^{#2}}}

\knowledgenewcommand{\Succ}{\mathit{\cmdkl{Succ}}}
\knowledgenewcommand{\St}[1]{\cmdkl{\S_{#1}}}

\knowledgenewcommand{\For}{\cmdkl{\F}}

\knowledgenewcommand{\guided}{\cmdkl{\rtimes}}

\knowledgenewcommand{\AGame}[2]{\cmdkl{\G(#1,#2)}}

\knowledgenewcommand{\rks}{\cmdkl{\mathtt{rk}^*}}
\knowledgenewcommand{\rk}{\cmdkl{\mathtt{rk}}}

\knowledgenewcommand{\Rg}[2]{\cmdkl{\R(#1,#2)}}
\knowledgenewcommand{\g}{\cmdkl{\mathtt{g}}}
\knowledgenewcommand{\RgS}{\cmdkl{\mathtt{Reg}}}

\title{Using games and universal trees to characterise the nondeterministic index of tree languages} %TODO Please add

\titlerunning{Games, universal trees and the index of tree languages} %TODO optional, please use if title is longer than one line

\author{Olivier Idir}{Université Paris Cité, CNRS, IRIF, France}{olivier.idir@ens-lyon.org}{https://orcid.org/0009-0003-3848-8515}{}%TODO mandatory, please use full name; only 1 author per \author macro; first two parameters are mandatory, other parameters can be empty. Please provide at least the name of the affiliation and the country. The full address is optional. Use additional curly braces to indicate the correct name splitting when the last name consists of multiple name parts.

\author{Karoliina Lehtinen}{CNRS, Université Aix-Marseille, LIS, Marseille, France}{karoliina.lehtinen@lis-lab.fr}{https://orcid.org/0000-0003-1171-8790}{}

\authorrunning{O. Idir and K. Lehtinen} %TODO mandatory. First: Use abbreviated first/middle names. Second (only in severe cases): Use first author plus 'et al.'

\Copyright{Olivier Idir and Karoliina Lehtinen} %TODO mandatory, please use full first names. LIPIcs license is "CC-BY";  http://creativecommons.org/licenses/by/3.0/


 %TODO mandatory; please add comma-separated list of keywords

\acknowledgements{We are grateful to Marcin Jurdzi\'nski for insightful discussions about attractor decompositions at Dagtuhl Seminar 24231, and of course to Dagstuhl for enabling such discussions.}

\relatedversion{This is a full version of an article of the same name published at ICALP 2025}

\category{Track B: Automata, Logic, Semantics, and Theory of Programming}

\EventEditors{Keren Censor-Hillel, Fabrizio Grandoni, Joel Ouaknine, and Gabriele Puppis}
\EventNoEds{4}
\EventLongTitle{52nd International Colloquium on Automata, Languages, and Programming (ICALP 2025)}
\EventShortTitle{ICALP 2025}
\EventAcronym{ICALP}
\EventYear{2025}
\EventDate{July 8--11, 2025}
\EventLocation{Aarhus, Denmark}
\EventLogo{}
\SeriesVolume{334}
\ArticleNo{159}
\begin{document}
	
	\maketitle
	
	\begin{abstract}
	The parity index problem of tree automata asks, given a regular tree language
$L$ and a set of priorities $J$, is $L$ $J$-feasible, that is, recognised by a nondeterministic parity automaton with priorities $J$? This is a long-standing open problem, of which only a few sub-cases
and variations are known to be decidable.
In a significant but technically difficult step, Colcombet and L\"oding reduced the problem to the uniform
universality of distance-parity automata.
In this article, we revisit the index problem using tools from the parity game literature.

We add some counters to Lehtinen's register game, originally used to solve parity games in quasipolynomial time,
and use this novel game to characterise $J$-feasibility. 
This provides a alternative proof to Colcombet and L\"oding's reduction.

We then provide a second characterisation, based on the notion of attractor decompositions and the complexity of their structure, as measured by a parameterised version of their Strahler number, which we call $n$-Strahler number. Finally, we rephrase this result using the notion of universal tree extended to automata: a guidable automaton recognises a $[1,2j]$-feasible language if and only if it admits a universal tree with $n$-Strahler number $j$, for some $n$.
In particular, a language recognised by a guidable automaton $\A$ is B\"uchi-feasible if and only if there is a uniform bound $n\in \NN$ such that all trees in the language admit an accepting run with an attractor decomposition of width bounded by $n$. Equivalently, the language is B\"uchi-feasible if and only if $\A$ admits a \textit{finite} universal tree.

While we do not solve the decidability of the index problem, our work makes the state-of-the-art more accessible and brings to light the deep relationships between the $J$-feasibility of a language and attractor decompositions, universal trees and Lehtinen's register game.  

	\end{abstract}
\section{Introduction}

Finite-state "automata" running on infinite structures are fundamental to the theory of verification and synthesis, where they model non-terminating systems. 
The complexity of an "automaton" is measured not only by the size of its state-space, but  also by the complexity of the acceptance condition. For instance, while the membership and non-emptiness questions for B\"uchi and coB\"uchi tree automata are in \ptime, for parity automata they are fixed-parameter tractable in the number of priorities in the parity condition, called its index~\cite{Calude2017DecidingPG}. In the modal $\mu$-calculus, the logic corresponding to parity tree automata, the alternation depth of a formula -- that is, the nesting depth of alternating least and greatest fixpoints -- coincides with the index of the corresponding parity automaton.

While for nondeterministic automata over infinite words, the B\"uchi acceptance condition suffices to recognise all $\omega$-regular languages~\cite{Rabin1968DecidabilityOS}, the classes of languages recognised by parity tree automata of each index form an infinite hierarchy, often called the parity, Mostowski, or Rabin-Mostowski index hierarchy. In other words, no fixed parity index suffices to recognise all $\omega$-regular tree languages, and this is the case for both nondeterministic~\cite{Niwinski1986OnFC}, and alternating~\cite{Bra98, Len96} tree automata.
A language is said to be $J$-feasible if it is recognised by a nondeterministic parity automaton of index $J$.
The nondeterministic index of an $\omega$-regular tree language is the minimal index $J$ for which it is $J$-feasible. The decidability of the index of a language %is classical problem in automata theory
%, often also referred to as the Mostowski or Rabin-Mostowski index problem, 
%that 
is one of the major open problems in automata theory.

In the case of infinite words, the \textit{deterministic} index of a language is decidable in \ptime~\cite{Niwinski1998RelatingHO}. In the case of infinite trees, however, not much is known. For languages given by deterministic parity automata, deciding their nondeterministic index is decidable~\cite{NW05}. Similarly, deciding if a language is recognisable with a safety/reachibility condition can be done in EXPTIME \cite{Walukiewicz2002DecidingLL}. CoB\"uchi-feasibility, as well as the  weak feasability of B\"uchi languages, are also decidable~\cite{Colcombet2013DecidingTW,SW16}. For the restricted class of game automata (which can be seen as the closure of deterministic automata under complementation and composition), the nondeterministic and alternating index problems are decidable~\cite{GameAutomata}. The most recent advance on the topic is that the \textit{guidable} index of a language is decidable~\cite{NS21}, where "guidable" automata, introduced by Colcombet and L\"oding~\cite{Guidable}, restrict the nondeterminism of the automaton without the loss of expressivity imposed by determinism.

The general nondeterministic index problem remains wide open. However, in a significant step, in 2008, Colcombet and Löding~\cite{Guidable} reduced the index problem of a tree language to the uniform universality of distance-parity automata. 
This remarkable result is, however, quite technical. In this article we present a similar result, (from which Colcombet and Löding's result can be obtained as a corollary, see~\cref{rmk:colcombet-loding}), using variations of known tools from the parity game literature -- namely, attractor decompositions, universal trees, the register index of parity games, and Strahler numbers. These are all notions that (re-)emerged in the aftermath of Calude et al.'s first quasipolynomial algorithm for solving parity games~\cite{Calude2017DecidingPG} to provide clarity on the newly established complexity bound. Here, we demonstrate that these tools also provide insight into the index hierarchy by using them to reformulate Colcombet and Löding's result and give an alternative proof.

Let us discuss each of these notions in more detail, in order to state our results.\\

\subparagraph*{The register index of parity games and $J$-feasibility.} Parity games are infinite two-player games in which two players, Adam and Eve, take turns moving a token along the edges of a graph labelled with integer priorities. Eve's goal is to ensure that the infinite path taken by the token satisfies the parity condition, that is, that the highest priority occuring infinitely often along the path is even. The acceptance of a tree by a parity tree automaton is determined by whether Eve wins a parity game based on the input tree and the automaton. 

In this article, we use the data-structure introduced in Lehtinen's quasipolynomial parity game algorithm~\cite{RegisterGames}. Lehtinen reduces solving a parity game to solving a new game, in which Eve must map the original game's priorities into a smaller priority range using a purpose-built data-structure, while guaranteeing that the sequence of outputs in this smaller range still satisfies the parity condition. Lehtinen shows that for a parity game of size $n$, Eve wins if and only if she also wins this new game with output range $O(\log n)$, which can be solved in quasipolynomial time.

Here we extend this game to the acceptance parity games of nondeterministic parity tree automata, that is, parity games with  unbounded or even infinite arenas. We furthermore add some counters (inspired by the Colcombet and L\"oding construction), which give  Eve some additional (but bounded) leeway in her mapping. We obtain a game that we call the parity transduction game $\Reg{J}{n}(G)$, played over a parity game $G$, parameterised by the output priority range $J$ , and the bound $n$ on the counters.

Our first contribution is showing that the $J$-feasibility of the language of a guidable automaton $\A$ (and we can always make the input guidable) is characterised by the existence of an integer $n$ such that the parity transduction game with parameters $J$ and $n$ coincides with the acceptance game of $\A$, written $\AGame{\A}{t}$ for an input tree $t$. In other words, a language is $J$-feasible whenever there is a uniform parameter $n$, such that whenever Eve wins the acceptance game $\AGame{\A}{t}$, she also wins the transduction game over it, with output range $J$ and parameter $n$.

\begin{restatable}{theorem}{thmfeasibilitygame}\label{thm:feasible-register}
Given a "guidable automaton" $\A$, $J$ an "index", the following are equivalent:
\begin{itemize}
\item $\Lang(\A)$ is $J$-"feasible".
\item There exists $n\in \NN$ such that for all $\Sigma$-tree $t$, $t\in \Lang(A)$ if and only if Eve wins $\Reg{J}{n}(\AGame{\A}{t})$.
\end{itemize}

\end{restatable}
 
 This corresponds to our version of the Colcombet-L\"oding reduction.
 We then proceed to reinterpret this characterisation in terms of \textit{attractor decompositions} and \textit{universal trees}.\\

\subparagraph*{Attractor decompositions}
describe the structure of Eve's winning strategies in a parity game, or, equivalently, of accepting runs of a parity tree automaton.
While this notion appears at least implicitly in many seminal works, %~\cite{Zielonka1998InfiniteGO,KV97,Klarlund1991ProgressMF}, 
e.g. Zielonka's algorithm for parity games \cite{Zielonka1998InfiniteGO}, Kupferman and Vardi's automata transformations~\cite{KV98} and Klarlund's~\cite{Klarlund1991ProgressMF} proof of Rabin's complementation theorem, 
it has more recently been explicitly studied  for mean payoff parity games~\cite{Daviaud_2018} and  parity games \cite{Attractor_decomposition_next,jurdzinski2022universalalgorithmsparitygames}.

While similar in spirit and structure to progress-measures~\cite{Jurdzinski2017SuccinctPM}, which count the number of odd priorities that might occur before a higher priority, attractor decompositions are more suitable for parity games on infinite arenas, where Eve might see an unbounded number of odd priorities in a row, as long as she is advancing in the attractor of some larger even priority. While progress measures, bounded by the size of a finite game, can be seen as a way to reduce parity games to \textit{safety} games, here we use attractor decompositions with bounded structure to reduce the priority range of the parity condition.
Like progress-measures, attractor decompositions have a tree-like structure, where the play only moves to the right if a suitably high even priorities occurs. The structure of these trees turns out to be closely tied to the index of a language.\\

\subparagraph*{{$n$-Strahler number}}
The Strahler number of a tree $t$ consists in the largest $h$ such that $t$ admits a complete binary tree of height $h$ as a minor.
Daviaud, Jurdzi\'nski and Thejaswini \cite{StrahlerNumber} proved an equivalence between the output range that Eve needs in Lehtinen's  game, called the game's register index, and the Strahler-number of the attractor decompositions of Eve's strategies.
 Inspired by this, we define, for $n\in \NN$, the $n$-Strahler number of a tree $t$, that consists in the largest $h$ such that $t$ admits a complete $(n+1)$-ary tree of height $h$ as a minor (by subtree deletion and single-child contraction; we do not allow edge contraction in the presence of siblings). The Strahler number corresponds to our $1$-Strahler number.
 Our second characterisation of the index of a languages is based on the $n$-Strahler number of attractor decompositions.

\begin{restatable}{theorem}{thmparityattractors}\label{cl:parity-attractors}
Given a "guidable" "nondeterministic parity tree automaton" $\A$, the following are equivalent:
\begin{itemize}
\item $L(\A)$ is $[1,2j]$-"feasible".
\item There is an $n\in \NN$ such that for all $t\in L(\A)$ there exists a "run" of $\A$ on $t$ with an "attractor decomposition" of "$n$-Strahler number" at most $j$.
\end{itemize}

\end{restatable}

In particular, B\"uchi feasibility coincides with the existence of a uniform bound on the width (i.e branching degree) of attractor decompositions needed by Eve.
Finally, we restate this result in terms of \textit{universal trees}, extended to automata, as follows.\\

\subparagraph*{{Universal trees}}
Given a set $\T$ of ordered trees of bounded depth, a tree $U$ is said to be universal for $\T$ if all $t\in \T$ can be obtained from $U$ by removing subtrees. We then say that $t$ is isomorphically embedded in $U$. 
This elegant notion emerged in the analysis of quasi-polynomial time parity game algorithms, as a unifying combinatorial structure that can be extracted from the different algorithms \cite{Czerwinski2018UniversalTG}.

We say that an ordered tree $U$ is universal \textit{for an automaton $\A$} if for all regular trees in the language of $\A$, there exists an accepting run with an attractor decomposition (seen as a tree) that can be isomorphically embedded in $U$.

 Then, the  $[1,2j]$-feasibility of the language of a guidable automaton $\A$ is characterised by the existence of an  ordered tree  universal for $\A$ of $n$-Strahler $j$, for some $n\in \NN$. B\"uchi-feasibility is equivalent to the existence of a \textit{finite} universal tree for $\A$.

\begin{restatable}{theorem}{thmuniversaltrees}\label{cl:universal-trees}
Given a "guidable" "nondeterministic parity tree automaton" $\A$, the following are equivalent:
\begin{itemize}
\item $L(\A)$ is $[1,2j]$-"feasible".
\item There exists an $n\in \NN$ and a tree $\U$ of "$n$-Strahler number" at most $j$  that is "universal@@automaton-univ" for $\A$.
\end{itemize}

\end{restatable}

\vspace{5mm}

While our work does not give us the decidability of the index problem, it provides new tools for tackling it and makes the state-of-the-art more accessible by relating it to other familiar concepts. We hope that the deep link between the index of a language and the structure of attractor decompositions will be helpful for future work. The remarkably simple characterisation of B\"uchi feasible languages, as those with attractor decompositions of bounded width, or, equivalently a finite universal tree, is particularly encouraging, as deciding B\"uchi-feasibillity is the next challenge for advancing on the index problem.

\ificalp
A full version of this article, with all the missing proofs, can be found on Arxiv \cite{fullversion}.
\fi

\section{Preliminaries}\label{sec:definitions}
% !TeX spellcheck = en_US
% !TEX root =  main.tex

The set of natural numbers $\{0, 1, \dots\}$ is denoted $\NN$, the set of strictly positive numbers is denoted $\NN^+$. The disjoint union of two sets $A$ and $B$ is denoted $A \sqcup B$. 
An ""alphabet"" is a finite non-empty set $\Sigma$ of elements, called letters.
$\Sigma^*$ and $\Sigma^\omega$ denote the sets ot finite and infinite ""words"" over $\Sigma$, respectively.
For $u$ a (possibly infinite) "word" and $n\in \NN$, the "word" $u|_{n}$ consists of the first $n$ letters of $u$. For $u$ and $v$ finite "words", $u \intro*\concat v$ denotes the concatenation of $u$ and $v$.
The length of a finite "word" $u$ is written $|u|$.

\AP An ""index"" $[i,j]$ is a non-empty finite range of natural numbers $I = \{i, i+1,\dots, j\} \subseteq \NN$.
Elements $c \in I$ are called ""priorities"". We say that an infinite sequence of "priorities" $(c_n)_{n\in \NN}$ is ""parity accepting"" (or simply \reintro*accepting) if $\limsup_{n \to \infty} c_n \equiv 0 \mod 2$, else it is "parity rejecting" (or "rejecting").

\subsection{Parity games}
\AP For $I$ an "index", $(V,E)$ a graph with $V$ a countable set of vertices and $L:E\to I$ an edge labeling, we call $G = (V,E,L)$ a ""$I$-graph"", or a "parity graph". We work with graphs in which every vertex has at least one successor. 
A graph (or tree) is said ""finitely branching"" if all its vertices have a finite number of exiting edges.

A graph is said ""even"" if all its infinite paths are "parity accepting".
For $G=(V,E,L)$ a "parity graph" and $V' \subseteq V$, the graph $G \intro*\restrict V'$ is the subgraph restricted to the vertices in $V'$. Similarly, for $E' \subseteq E$, the graph $G \setminus E'$ corresponds to $(V,E\setminus E', L')$ with $L'$ the restriction of $L$ to $E\setminus E'$.

\AP Let $G = (V,E,L)$ a "parity graph", and $E' \subseteq E$. The ""attractor"" of $E'$ in $G$ is the set
$\intro*\attr(E',G) := \{v\in V|\forall \text{ infinite path } \rho \text{ from } v \text { in }G,\ \rho \text{ has an edge in } E'\}$. Similarly, if $V' \subseteq V$, we define its "attractor" as the set $\attr(V',G)$ of vertices from which all infinite paths eventually pass by $V'$. Note that $V' \subseteq \attr(V',G)$.

\AP A ""parity game"" played by players Eve and Adam consists in a "parity graph" $\G = (V,E,L)$ with a partition of $V$ in two sets: $V = V_E \sqcup V_A$, controlled respectively by Eve and Adam. % and with $E$ such that all vertices have an outgoing edge. -> Pas nécessaire, on a dit plus haut travailler avec des graphes sans sommet terminaux
A ""play"" of $\G$ starting in $v\in V$ consists in an infinite sequence of edges $\rho := (e_i)_{i\in \NN}$ forming an infinite path starting in $v$. 
A "play" $(e_i)_{i\in \NN}$ is ""winning"" for Eve (or simply "winning") if $(L(e_i))_{i\in \NN}$ is "parity accepting", else it is said to be "losing" (for Eve, and "winning" for Adam).

\AP A ""strategy"" for Eve consists of a function $\sigma : E^* \to E$ such that, for all play $\rho$, for all $n \in \NN$, if $\rho_{|n}$ ends in a vertex $v \in V_E$, $\sigma(\rho_{|n})$ is an edge from $v$. A "play" $\rho$ is said to be ""consistent with"" the strategy $\sigma$ if for all $n$, $\rho_{|n}$ ending in a vertex of $V_E$ implies that $\rho_{|n+1} = \rho_{|n}\sigma(\rho_{|n})$. 
We say that a Eve "strategy" $\sigma$ is ""winning@winning strategy"" from vertex $v\in V$ if all plays "consistent with" $\sigma$ starting in $v$ are "winning". We similarly define "strategies" for Adam, winning when all plays "consistent with" them are "winning" for Adam.

\AP Parity games enjoy positional ""determinacy"": one of the players always wins with a strategy that only depends on the current position~\cite{EJ91}.

A "strategy" for Eve in a game $\G = (V_E\sqcup V_A,E,L)$ induces an Adam-only "game" $\G'$ played on the unfolding of $\G$, from which are removed all the edges that Eve does not choose. This game can be seen as a "parity graph", as the partition of the vertex set is now a trivial one, and it is "even" if and only if Eve's strategy is winning.

\subsection{Attractor decomposition}

An "attractor decomposition" of an "even" "parity graph" $G$ is a recursive partitionning of $G$. The intuition is that it identifies subgames of $G$ in which the top priorities $h$ (even) and $h-1$ (odd) do not occur and orders them so that a path must always eventually either stay within a subgame (and never see $h-1$ again), advance in the order (potentially seeing $h-1$ finitely many times in between by advancing through the attractor of a subgame), or see the higher even priority $h$. Each subgame is then decomposed recursively, with respect to the priority $h-2$. As the number of subgames is countable, such a decomposition witnesses that the "parity graph" is indeed "even".
An "attractor decomposition" has a tree-like structure, induced by the order on the subgames (which corresponds to the order of sibling nodes), and their sub-decompositions.

\AP Given a "parity graph" $G=(V,E,L)$ with maximal priority at most some even $h$, and $\kappa$ an ordinal, a ($h$-""level"", $\kappa$-""width"")-""attractor decomposition"" of $G$, if it exists, is recursively defined to be $D=(H,A_0,\{(S_i,A_i,D_i)\}_{0<i<\ell})$ where:
\begin{itemize}
	\item $\ell\leq \kappa$,
	\item $H\subseteq E$ is the set of edges in $G$ of priority $h$,
	\item $A_0= \attr(H,G)$,
	\item For every $0<i<\ell$, let $V_i=V\setminus \bigcup_{j< i}A_j$ and $G_i=(G\setminus H)\restrict V_i$. Then:
	\begin{itemize}
		\item $S_i\subseteq V_i$  is non-empty, such that $(G\setminus H)\restrict S_i$ only contains edges with priorities up to $h{-}2$, has no terminal vertices and is closed under successors in $G_i$,
		\item $A_i$ is  $\attr(S_i,G_i)$,
		\item $D_i$ is a ($(h{-}2)$-"level", $\kappa$-"width")-"attractor decomposition" of $(G\setminus H) \restrict S_i$,
	\end{itemize}
	\item $V=\bigcup_{i<\ell} A_i$,
	\item A ($0$-height, $\kappa$-"width")- and a ($h$-"level", $0$-"width")-"attractor decomposition" is just $(H,V)$: the entire graph is in the attractor of the edges of highest priority.
\end{itemize}

\begin{restatable}{lemma}{lemADreachability}\label{cl:ADreachability}	
	Given a "parity graph" $G$ that admits an "attractor decomposition"\\
	$(H, A_0,\{(S_i,A_i,D_i)\}_{0<i < \kappa})$, the set $A_j$ is unreachable from $A_i$ in $G\restrict \bigcup_{0<\ell < \kappa} A_{\ell}$ for all $i$ and $j$ such that $0<i<j < \kappa$.
\end{restatable}

\begin{proof}
	We proceed by transfinite induction on $i$. $A_j$ is unreachable from $A_1$ in $G\restrict \bigcup_{0<\ell<\kappa} A_{\ell}=G\restrict V_1$ since all paths from $A_1$ lead to $S_1$, which is closed under successors in $G\restrict V_1$ by definition and $A_j$ is disjoint from $A_1$ (by definition since it is an "attractor" in $G\restrict V_j$, a graph disjoint from $A_1$ for $j>1$).

	For the induction step, assume $A_j$ is unreachable in $G\restrict V_1$ from all $A_{\ell}$ for $0<\ell< i$. Since $A_i$ is by definition an "attractor" of $S_i$ in $G\restrict V_i$, any path from $A_i$ in $G\restrict V_i$ ends up in $S_i$ without leaving $A_i$, and then $S_i$ is closed under successors in $G\restrict V_i$. Therefore, all paths from $A_i$ in $G\restrict V_i$ can only exit $A_i$ by entering $\bigcup_{0<\ell<i}A_{\ell}$. Then, from the induction hypothesis, such a path cannot reach $A_j$.\qedhere
	
\end{proof}

\begin{restatable}{lemma}{evenIffDecomposition}\label{cl:EvenIffDecomposition}
	
	A "parity graph" is "even" if and only if it admits an "attractor decomposition".

\end{restatable}

\begin{proof}
	If a "parity graph" $G$, of maximal priority at most some even $h$, is "even", we can construct an "attractor decomposition" recursively for it as follows.
	Let $H$ be the set of edges of priority $h$ and $A_0$ the "attractor" of $H$. 
	
	Then, we define each $S_i$ and $A_i$ for $i>0$ inductively. First let $V_i=V\setminus \bigcup_{\ell< i} A_\ell$ for $i\in \NN$  or an ordinal.
	$V_i$ is either empty, or $G_i=G\restrict V_i$ is an "even" "parity graph" with maximal priority no larger than $h{-}1$. Let $S_i$ consist of all positions of $G_i$ from where $h{-}1$ can not be reached. That is, $S_i$ is "even" (being a subgraph of $G$) and only has edges of priority up to $h-2$
	If $V_i$ is non-empty, there must be such positions, since otherwise one could build a path which sees infinitely many $h{-}1$, contradicting that $G_i$ is an "even graph". Let $A_i$ be the "attractor" of $S_i$ in $G_i$ and let $D_i$ be an "attractor decomposition" of level $h{-}2$ of $S_i$, which we can exhibit by recursion.

	Let us assume $V_i$ is non-empty for all countable ordinals $i$. Since all the $V_i$s are disjoint, their union is uncountable, since there are uncountably many ordinals smaller than $\omega_1$. However we only work with countable graphs, which gives a contradiction. Thus $V_i$ must be empty for some countable ordinal $i$. Hence $V=\bigcup_{\ell <i}A_i$.

	For the other direction, assume that a "parity graph" has an "attractor decomposition" $(T, A_0,\{(S_i,A_i,D_i)\}_{0<i<\kappa})$ of "level" $h$.
	
	We proceed by induction on the "level" of the "attractor decomposition", that is, the number of priorities in $G$. The base case of the unique priority $0$ is trivial since all paths are "parity accepting".
	
	For the induction step, we observe that any infinite path of $G$ must either eventually remain in some $S_\ell$ for $1\leq \ell< \kappa$ or reach $H$ infinitely often, due to \cref{cl:ADreachability}. In the latter case, the path is "parity accepting" since all edges in $H$ are of the highest possible priority. In the former case, since each $S_\ell$ has an attractor decomposition of height $h{-}2$, by the induction hypothesis, we are done. \qedhere
	
\end{proof}

\subsection{$\Sigma$-trees and automata}
\AP A ""$\Sigma$-tree"" (or just \reintro*{tree}) is a function $t : \{0,1\}^* \to \Sigma$. The set of all $\Sigma$-trees is denoted ${\Tr {\Sigma}}$. A tree is ""regular"" if it is finitely representable, that is, if it is the unfolding of a rooted graph. We denote $\intro*\RgS_\Sigma$ the set of "regular" trees of $\Tr{\Sigma}$.

\AP An infinite word $b \in \{0,1\}^\omega$ is called a ""branch"".
Given a  "tree" $t\in \Tr\Sigma$, a ""path@@tree-p"" $p$ (along a "branch" $b$) is a sequence $(p_i)_{i \in \NN} := (t(b_{|i}))_{i\in \NN}$.

\AP A ""nondeterministic $I$-parity tree automaton"" (also called $I$-automaton, or "automaton" of "index" $I$) is
a tuple $A = (\Sigma, Q_A , q_{i,A}, \Delta_A, \Omega_A )$, where $\Sigma$ is an alphabet, $Q_A$ a finite set of ""states"", $q_{i,A} \in Q_A$ an initial "state", $\Delta_A \subseteq Q_A \times \Sigma \times Q_A \times Q_A$ a transition relation; and $\Omega_A : \Delta_A \to I^2$ a "priority" mapping over the edges. 
A ""transition"" $(q, a, q_0 , q_1) \in \Delta_A$, is said to be from the "state" $q$ and over the letter $a$. By default, all "automata" in consideration are complete, that is, for each state $q \in Q_A$ and letter $a\in \Sigma$, there is at least one transition from $q$ over $a$ in $\Delta_A$.
When an automaton A is known from the context, we skip the subscript and write just $Q,\Delta$, etc.

\AP For $q,q' \in Q$, a ""path@@automaton-p"" from $q$ to $q'$ is  a finite "transition" sequence $(q_j, a_j, q_{j,0}, q_{j,1})_{j< N} \in \Delta^N$ such that $q = q_0$, and $\forall j < N, q_{j+1}\in \{q_{j,0}, q_{j,1}\}$ with $q_{j+1} = q'$.

\AP A tree is said to be "accepted" by an "automaton" $A$ if Eve wins a game defined by the product of this tree and the automaton, in which Eve chooses the transitions in $A$ and Adam chooses the direction in $t$. More formally, given a tree $t \in \Tr\Sigma$, and an $I$-"automaton" $A$, the ""acceptance game"" of $A$ on $t$, also denoted $\intro*\AGame{A}{t}$, is the "parity game" obtained by taking the product of $A$ and $t$. Its arena consists in $\{0,1\}^*\times (Q_A \cup \Delta_A)$, where all the positions of the shape $\{0,1\}^*\times Q_A$ are controlled by Eve, and the others by Adam. 
\begin{itemize}
	\item When in a position $(w,q) \in \{0,1\}^* \times Q_A$, Eve chooses a transition $e\in \Delta_A$ of the shape $(q,t(w),q_0,q_1)$, and the play proceeds to the state $(w,e)$. All these transitions have for label the minimal priority in $I$.
	\item Let $q\in Q_A$ and $e=(q,a,q_0,q_1) \in \Delta_A$. In a position $(w,e)$, Adam chooses either $0$ or $1$, and the games then moves towards either $(w\concat 0, q_0)$ or $(w\concat 1, q_1)$. For $\Omega_A(e) = (i_0,i_1)$, these transitions have priorities $i_0$ and $i_1$, respectively.
\end{itemize}
\AP We say that $t$ is ""accepted@@tree"" by $A$ if Eve wins $\AGame{A}{t}$. The set of "trees" "accepted" by $A$ is called the ""language"" of $A$ and is denoted $\intro*\Lang(A)$. We say that $\A$ ""recognizes"" $\Lang(A)$.

\AP If we fix a "strategy" for Eve, the "acceptance game" becomes an Adam-only game, called a ""run"" of $A$ on $t$. We observe that it is played on a "parity graph" in the shape of a binary tree. We thus observe that a "run" can be considered as a tree in $\Tr{\Delta_A}$. This "run" is won by Adam if and only if there exists a "parity rejecting" "branch". In this case, it called a ""rejecting run"", else it is an "accepting run".

If $A$ is an "$I$-automaton", such a "run" over $t$ induces an $I$-labelling of $t$, which, for convenience, we consider to be on edges.

%can also be considered as a tree in $\Tr I$, up to forgetting the exact edges (and neglecting the first priority, as the priorities are now witnessed on the nodes. We can neglect it as the parity condition is prefix-independent). This tree is "even" if and only if the run is "accepting".\karoliina{Add distinction between labelling and seeing a run as a tree}

\AP A set of trees $L \subseteq \Tr{\Sigma}$ is an ""$\omega$-regular tree language"" if it is of the form $\Lang(A)$ for some "automaton" $A$. It is said  to be ""$I$-feasible"" if furthermore $A$ is of "index" $I$.

\subsection{Guidable automata}

The notion of a "guidable automata" was first introduced in \cite{Guidable}.
%They can be conceived as a form of history-determinism applied to tree automata. Tree automata are not necessarily determinisable: for instance, the language of trees that have an \texttt{a} somewhere is a regular language, recognized by a simple two-state nondeterministic reachability\oi{or Büchi, as we did not introduce reachability condition} automaton. Yet there does not exist a deterministic automaton recognizing this language, as it would need to know in advance in which direction to look for the \texttt{a}.\\
%"Guidability" circumvents this issue: instead of guessing directions, a "guidable automaton" takes as input an accepting run of another automaton recognizing a sublanguage, and uses it to solve its nondeterminism. 
Intuitively, they are "automata" that fairly simulates all language equivalent "automata". "Guidable automata" are fully expressive \cite[Theorem 1]{Guidable} and are more manageable than general nondeterministic automata. 

\AP Fix two "automata" $A$ and $B$ over the same "alphabet" $\Sigma$. A ""guiding function""
from $B$ to $A$ is a function $g : Q_A \times \Delta_B \to \Delta_A$ such that $g(p, (q, a, q_0 , q_1 )) = (p, a, p_0 , p_1)$
for some $p_0, p_1 \in Q_A$ (i.e. the function g is compatible with the "state" $p$ and the letter $a$). 

If $\rho \in Tr_{\Delta_B}$ is a "run" of $B$ over a "tree" $t \in \Tr\Sigma$ then we define the "run" $g (\rho) \in Tr_{\Delta_A}$ as follows. We define inductively $q : \{0,1\}^* \to Q_A$ in the following fashion: $q(\varepsilon) = q_{i,A}$, and supposing $q(u)$ to be defined, for $g(q(w),\rho(w))=(q(w),t(w),q_0,q_1)$, we let $q(u\concat 0),q(u\concat 1)$ to be respectively $q_0,q_1$. We can then define the run $g (\rho) \in \Tr{\Delta_A}$ as 
$$g(\rho): u \mapsto g(q(u),\rho(u)).$$
Notice that directly by the definition, the "tree" $g(\rho)$ is a "run" of $A$ over $t$. 

\AP We say that a "guiding function" $g : Q_A \times \Delta_B \to \Delta_A$ ""preserves acceptance"" if whenever $\rho$ is an "accepting run" of $B$ then $g(\rho)$ is an "accepting run" of $A$. We say that an "automaton" $B$ ""guides"" an "automaton" $A$ if there exists a "guiding function" $g : Q_A \times \Delta_B \to \Delta_A$ which "preserves acceptance". In particular, it implies that $\Lang(B) \subseteq \Lang(A)$.

\AP An "automaton" $A$ is ""guidable"" if it can be "guided by" any "automaton" $B$ such that $L(B) = L(A)$
(in fact one can equivalently require that $L(B) \subseteq L(A)$, see \cite[Remark 4.5]{lodingHDR}).
We will use the following fundamental theorem, stating that "guidable automata" are as expressive as "non-deterministic ones@automata".

\begin{theorem}[{\cite[Theorem 1]{Guidable}}]
	For every regular tree language $L$, there exists a guidable automaton recognizing $L$. Moreover, such an automaton can be effectively constructed from any non-deterministic automaton for $L$.
\end{theorem}

 \subsection{Ordered trees}

 \AP We define inductively ""ordered trees"" of finite "depth". 
 They are either the ""leaf"" tree $\langle \rangle$ of "depth" $\intro*\d(\langle\rangle) = 1$, or a tree $T = \langle (T_k)_{k\in K} \rangle$ 
 where $\forall k, T_k$ is an ordered tree of finite "depth", and $K$ is a well-ordered countable set. 
 The ""children"", ""siblings"" and ""subtree"" relation $\intro*\subt$ are defined in the usual way.
 We denote $\prec$ the order relation between the "siblings" of a tree $\langle (T_k)_{k\in K} \rangle$. That is, for $k,k' \in K$, we have $T_k \prec T_{k'}$ when $k < k'$ for $<$ the well-order of $K$.  By abuse of notation, we say that $T_1 \prec T_2$ if $T_1\subt T'_1, T_2 \subt T'_2$ and $T'_1 \prec T'_2$.
 
 \AP From their definitions, it is clear that "attractor decompositions" are tree-shaped. To make this explicit, the ""tree-shape"" of an "attractor decomposition" $D= (H,A_0,\{(S_i,A_i,D_i)\}_{0<i<\kappa})$ is defined inductively as $\langle\rangle$ if $\kappa= 0$, else, defining $(T_i)_{0<i<\kappa}$ the "tree-shapes" of the $(D_i)_{0<i<\kappa}$, $D$ has "tree-shape" $\langle (T_i)_{0<i<\kappa}\rangle$.
 Observe that the "width" of an "attractor decomposition" corresponds to an upper-bound on the branching degree of its "tree-shape".
 
 \AP We extend the notion of "tree-shape" to "runs": a run has "tree-shape" $t$ if it has an "attractor decomposition" of "tree-shape" $t$.
 
\AP We say that an "ordered tree" $T = \langle (T_i)_{i\in I}\rangle$ is ""isomorphically embedded"" in a tree $T' = \langle (T'_j)_{j\in J}\rangle$ if either $I$ is empty, or of there exists $\phi: I \to J$, strictly increasing, such that $\forall i \in I$, $T_i$ is "isomorphically embedded" in $T'_{\phi(i)}$.
Intuitively, this implies the existence of a map from the "subtrees" of $T$ to the "subtrees" of $T'$, where the root of $T$ is mapped onto the root of $T'$, and the children of every node must be mapped injectively and in an order-preserving way onto the children of its image.

\AP Let $\T$ be a set of "ordered trees". We say that a tree $U$ is ""universal for $\T$"" if all the trees of $\T$ can be "isomorphically embedded" in $U$.

\section{Game characterisation of the parity index}\label{sec:register-game}
% !TEX root =  main.tex

In this section we define "priority transduction games", based on the register games from Lehtinen's algorithm  in~\cite{RegisterGames}, augmented with some counters. We characterise the "$J$-feasibility" of a "language" $\Lang(A)$, where $A$ is a "guidable automaton", by the existence of a uniform bound $n\in \NN$ such that a "tree" is in $\Lang(A)$ if and only Eve wins the $J$-"priority transduction game" on $\AGame{A}{t}$, with counters bounded by $n$.

The idea of these "priority transduction games" is that in addition to playing the "acceptance game" of an $I$-automaton $\A$ over a tree $t$, which has priorities in $I$, Eve must map these priorities on-the-fly into the "index" $J$. In the original games from~\cite{RegisterGames}, she does so by choosing at each turn a register among roughly $\frac{\lvert J \rvert }{2}$ registers. Each register stores the highest priority seen since the last time it was chosen. Then, the output is a priority in $J$ which depends on both the register chosen and the parity of the value stored in it. 
Here, the mechanism is similar, except that we additionally have counters that allow Eve to delay outputting odd priorities a bounded number of times.

Intuitively, the registers, which store the highest priority seen since the last time they were chosen, determine the magnitude of the output, while their content's parity decides the output's parity.
This allows Eve to strategically pick registers so that odd priorities get eclipsed by higher even priorities occuring soon after. However, a large odd priority occuring infinitely often will force Eve to produce odd outputs infinitely often.
The counters give Eve some error margin, whereby she can pick a register containing an odd value \textit{without} outputting an odd priority, up to $n$ times in a row. 

\AP Formally, for $J$ a priority "index" (of minimal value assumed to be $1$ or $2$ for convenience), $n \in \NN$, the ""$J,n$-priority transduction game"" is a game played by Eve and Adam, over an $I$-"parity graph" $G=(V,E,L)$ for $I$ an "index". It has two parameters, $J$ the output "index" and $n$ the bound of its "counters", and is denoted $\intro*{\Reg{J}{n}}(G)$. A configuration of the game corresponds to a position $p\in V$, a value in $I$ for each \reintro*"register" $r_j$ for even $2j\in J$ (if $1 \in J$, there is an additionnal "register" $r_0$), and a value between $0$ and $n$ for each \reintro*"counters" $c_{i,j}$ with $i$ odd $\in I, j$ such that $r_j$ is a "register".\\
Starting from some initial vertex $p_0\in V$ with "counters" set to $0$ and "registers" set to the maximal even priority in $I$, the game proceeds as follows at step $l$ :
\begin{itemize}
	\item Adam chooses an exiting edge $e = (p,p')\in E$ ; the position becomes $p'$.
	\item Eve chooses a "register" $r_j$.
	\item The game produces the output $w_l$ :
		\begin{itemize}
		\item if $j = 0$, $w_l=1$ (recall that $r_0$ is a "register" iff $1 \in J$). Else,
		\item if $r_j$ is even, $w_l=2j$.
		\item Else, if $c_{r_j,j} = n$, it is said to reach $n+1$ before being reset: $w_l=2j+1$ and $c_{r_j,j} := 0$. If $2j+1 \notin J$, Eve loses instantly.
		\item else, $w_l = 2j$ and $c_{r_j,j} := c_{r_j,j} +1$
		\end{itemize}
	\item If $L(e)$ is even, let $i:= L(e)$ be the label of the current edge, else Eve chooses an odd $i$ such that $L(e)\leq i$ (choice $\sharp$\label{sharp}). 
	Then the following updates occur : 
	\begin{itemize}
		\item Smaller "counters" are reset : $\forall i' < i, c_{i',j} := 0$ and $ \forall j'<j, c_{r_j,j'} := 0$,
		\item "Registers" get updated : $r_j := i$ and $\forall j' > j$, $ r_{j'} := \max(i, r_{j'}) $, 
	\end{itemize} 
\end{itemize}
Eve wins if the infinite sequence of outputs $(w_l)_{l\in \NN}$ is "parity accepting", else Adam wins.\\

In order to explain two aspects of this game that were not covered in the initial intuition: the minimal "register" $r_0$ allows Eve to wait, for a finite but unbounded time, for better priorities to override the register contents. It thus corresponds to outputting a minimal odd priority.
The choice $\sharp$ allows her to break a sequence dominated by many identical odd priorities $i$ in a sequence with some greater odd $i'$ in between, resetting the "counters" albeit at the cost of witnessing a greater odd priority.\footnote{This is a technical adjustement that is convenient in the proof of~\cref{lem:Strahler-to-Reg}}

Let $\G$ be a "parity game" of "index" $I$, $n\in \NN$, $J$ an "index". We define the game $\Reg{J}{n}(\G)$ as the game $\Reg{J}{n}$ where, instead of following a path of "parity graph" chosen by Adam, it follows an ongoing play of $\G$ where the player owning the current position $q$ chooses its move in $\G$ at each step, before Eve chooses her "register". It corresponds to the composition of $\Reg{J}{n}$ with the game $\G$. If we fix a strategy $\sigma$ for Eve in $\G$, we observe that $\Reg{J}{n}(\G)$ corresponds exactly to the "priority transduction game" $\Reg{J}{n}$ over the Adam-only game $\G_\sigma$ induced by $\sigma$ in $\G$, and that Eve wins $\Reg{J}{n}(\G)$ if and only if she wins $\Reg{J}{n}(\G_\sigma)$.

Note that $\Reg{J}{n}(\G)$ is a parity game, and therefore "determined".

We show that this transduction game characterises the "index" of a "regular tree language".

\thmfeasibilitygame*

%\begin{theorem}\label{thm:feasible-register}
%Given a "guidable automaton" $\A$, $J$ an index, the following are equivalent:
%\begin{itemize}
%\item $\Lang(\A)$ is $J$-"feasible".
%\item There exists $n\in \NN$ such that for all $\Sigma$-tree $t$, $t\in \Lang(A)$ if and only if Eve wins $\Reg{J}{n}(\AGame{\A}{t})$
%\end{itemize}
%\end{theorem}

For the upward implication, it suffices to observe that the transduction game is captured by a finite state $J$-automaton describing the register contents and counter values (bounded by $n$), with nondeterministic choices corresponding to Eve's choices, and a $J$-parity condition corresponding to the outputs. Then, the $J$-automaton equivalent to $A$ is the composition of $A$ with this $J$-automaton. 
\ificalp
The details, which are as one would expect, are in the full version.
\else
The details, which are as one would expect, are in the appendix.
\fi
\begin{restatable}{lemma}{automataComposition}\label{lem:automata-composition}
	Let $J$ be a priority "index" and let $A$ be a "guidable automaton" such that there exists $n\in \NN$ such that for all $\Sigma$-tree $t$, $t\in \Lang(A)$ if and only if Eve wins $\Reg{J}{n}(\AGame{A}{t})$. Then there exists an "automaton" of "index" $J$ such that $\Lang(B)=\Lang(A)$.
\end{restatable}

The rest of the section focuses on the downward implication of~\cref{thm:feasible-register}. We first show that Eve can only win $\Reg{J}{n}(G)$ for $G$ an "even" "parity graph", which implies that Eve loses $\Reg{J}{n}(\AGame{A}{t})$ for any $t\notin \Lang(A)$.

\begin{restatable}{lemma}{regRejectsRejecting}\label{lem:reg-rejects-rejecting}
	Let $G$ a "parity graph". If $G$ is not "even", then for all $J,n$, Adam wins $\Reg{J}{n}(G)$
\end{restatable}

\begin{proof}[Proof sketch]
If the underlying play in the parity graph sees a maximal odd priority $i$ infinitely often, then the most significant register $r_j$ that Eve picks infinitely often contains $i$ infinitely often when  picked. The counter $c_{i,j}$, which is eventually never reset, reaches $n$ infinitely often, making the maximal output priority that occurs infinitely often odd.
\end{proof}

Then, it remains to show that if the language of guidable $A$ is $J$-feasible, then for some $n\in \NN$ Eve wins $\Reg{J}{n}(\G(A,t))$ for all $t\in \Lang(A)$.
To do so, we first analyse the relation between guided and guiding runs, and show that the preservation of global acceptance implies a more local version that resctricts differences in the parity of the dominant priority over long segments of both runs. We will then use this to show that Eve can win the transduction game by using a run of $A$ guided by an accepting run of an equivalent $J$-automaton and choosing registers corresponding to the priorities of the guiding run.

The following lemma, obtained by a simple pumping argument 
\ificalp
(see full version)
\else
(see Appendix)
\fi
 expresses that between all pumpable pairs of states, that is, pairs of states that are not distinguished by either run, if the guiding run is dominated by en even priority, then so is the guided one. 

\begin{restatable}{lemma}{pumping}\label{cl:pump}
	Let $A,B$ be "automata", let $t\in \Tr{\Sigma}$, let $\rho_A, \rho_B$ be "accepting runs" over $t$ of $A$ and $B$, respectively, where $\rho_A$ is "guided by" $\rho_B$. We consider these "runs" as trees in $\Tr{\Delta_A},\Tr{\Delta_B}$ respectively. 
	Given $u,v \in \{0,1\}^*, \{0,1\}^+$ such that $\rho_A(u) = \rho_A(u\concat v)$, and such that $\rho_B(u) = \rho_B(u\concat v)$, if the greatest priority encountered between positions $u$ and $u \concat v$ in $\rho_B$ is even, so is the greatest priority encountered in this segment in $\rho_A$.
\end{restatable}

\AP We now capture this relation with the notion of a labelling being $n$-bounded by the other. Let $L_I: E\rightarrow I$ and $L_J:E\rightarrow J$ be two labellings of a graph $G= (V,E)$ (or tree) and let $n \in \NN$. We say that $L_I$ is ""$n$-bound"" by $L_J$ if there is no finite path $\pi$ in $G$, segmented into consecutive paths $\pi_0,\pi_1,\dots \pi_n$ such that for some odd $i$ and some even $j$, the maximal priority on the $L_I$- and $L_J$-labels of each $\pi_m, m\in [0,n]$ are $i$ and $j$, respectively.

From~\cref{cl:pump} we obtain that the labelling induced by a guided run is $n$-bound by the one induced by its guide, with $n$ the product of the sizes of the two automata:

\begin{restatable}{lemma}{guidableNBound}\label{lem:guidable-n-bound}
	Let $A$ a guidable automaton. If $\Lang(A)$ is $J$-feasible witnessed by an "automaton" $B$, for $n := |A| |B|+1$, for all $\Sigma$-tree $t \in \Lang(A)$, for $\rho_A$ the "run" of $A$ on $t$ "guided by" an "accepting run" $\rho_B$ of $B$ over $t$, the labelling $L_A$ induced by $\rho_A$ is "$n$-bound" by the labelling $L_B$ induced by $\rho_B$.
\end{restatable}

We use "$n$-boundedness" to show that Eve can use a run $\rho_B$ of an equivalent $J$-"automaton" to choose her "registers" to win in $\Reg{J}{n+1}(\rho_A)$ for $\rho_A$ "accepting run" of $A$ guided by $\rho_B$.

\begin{lemma}\label{cl:bounded-to-strategy}
	Let $I,J$ be indices, $n \in \NN^+$, and $\rho_I:E\rightarrow I$ and $\rho_J:E\rightarrow J$ two "even" $I$- and $J$- labelling of the same graph $(E,V)$. If $\rho_I$ is "$n$-bound" by $\rho_J$, Eve wins $\Reg{J}{n+1}(\rho_I)$.
\end{lemma}
\begin{proof}
	We will describe a "winning strategy" $\sigma$ for Eve in $\Reg{J}{n}(\rho_I)$. We recall that at each step, she has two choices : the choice of register and the choice of some $i\in I$ (choice $\sharp$). We once again suppose for convenience that $\min(J)\in \{1,2\}$. After seeing an edge of priority $i'$ in $\rho_I$, Eve chooses $i:= i'$. Given priority $j'$ seen in $\rho_J$, for $j:= \lfloor \frac{j'}{2}\rfloor$, Eve chooses the "register" $r_j$. This strategy being fixed, let us verify that Eve wins in all plays "consistent with" $\sigma$. Let $b$ such a play.
	
	As $\rho_J$ is "even", its maximal infinitely recurring priority along $b$ is even; we denote it $2j^{*} \in J$ (and $j^* \neq 0$, as $0 \notin J$). We look past the position where we no longer see any priority superior to $2j^*$. Then, as $2j^*$ is infinitely recurring, we observe that infinitely often the output is $w_l = 2j^*$, as this is the default output when choosing the register $r_{j^*}$. Let us show that the game outputs at most $\lceil\frac{|I|}{2}\rceil$ times $2j^*+1$.
	
	If it were to output $2j^*+1$ more than $\lceil\frac{|I|}{2}\rceil$ times, then there would be some counter $c_{i,2j^*}$ that would reach $n+2$ twice, for some odd $i$. We look at the first time $t_0$ where this counter reaches $n+2$: after time $t_0$, the counter $c_{i,j^*}$ has value $0$. We look at the $n+2$ times at which $c_{i, 2j^*}$ is incremented after $t^*$, denoted $(t_l)_{l\in [1,n+2]}$, and note $\pi_l$ the path between $t_l$ and $t_{l+1}$. Along these paths, we encounter no priority greater than $i$ in $\rho_I$, nor greater than $j^*$ in $\rho_J$, as these would reset $c_{i,j^*}$ (except, possibly, at time $t_{n+2}$ where we can witness a $i'>i$ after the counter has reached $n+2$). Additionally, for $l\in [1,n+1]$, as $c_{i,j^*}$ is incremented at time $t_l$, then there is a $2j^*$ in $\rho_J$ at the edge preceding $t_l$ (as this priority cannot be $2j^*+1$) and $r_{j^*} = i$ at time $t_l$, after which $r_{j^*}$ becomes the priority just seen in $\rho_I$. Therefore, $\forall l \in [1,n+1]$ between $t_{l-1}$ and $t_l $, we see at least once a $2j^*$ in $\rho_J$, and at least once a $i$ in $\rho_I$ – and they thus dominate these segments. This contradicts that $\rho_I$ is "$n$-bound" by $\rho_J$; therefore the output $2j^*+1$ does not repeat more than $\lceil\frac{|I|}{2}\rceil$ times.
	
	If $2j^* = \max(J)$, we set $t_0 := 0$. The hypothesis used for the previous reasoning still hold – initially all counters have value 0 and there is no $2j^*+1$ in $\rho_J$ – so we obtain that there no counter $c_{i,j*}$ that reaches $n+2$, which prevents instant loss.
	Therefore, the play along $b$ is won by Eve, which concludes.
\end{proof}

\cref{lem:reg-rejects-rejecting} implies that Eve loses $\Reg{J}{n+1}(\G(A,t))$ for $t\notin \Lang(A)$. If $\Lang(A)$ is $J$-feasible as witnessed by a $J$-automaton $B$, then for all $t\in \Lang(A)$, from~\cref{lem:guidable-n-bound}, Eve has a run $\rho_A$ that is $n$-bounded by an accepting run of $B$, which, from~\cref{cl:bounded-to-strategy}, implies that Eve wins $\Reg{J}{n+1}(\G(A,t))$, concluding the proof of~\cref{thm:feasible-register}.

\begin{remark}\label{rmk:colcombet-loding}
To obtain Colcombet and L\"oding's result from ours, it suffices to encode the transduction game as a distance-parity automaton that on an input tree $t$ computes a bound $n$ on the counters such that Eve wins  $\Reg{J}{n}(\G(A,t))$. Then, like in~\cite[Lemma 3]{Guidable}, there is a distance-parity automaton that is uniformly universal if and only if $A$ is $J$-feasible.
\end{remark} 

% !TEX root =  mostowski_games_icalp.tex

\knowledgenewcommand{\Sn}[2]{\cmdkl{\mathcal{S}_{#1}(#2)}}
\knowledgenewcommand{\Fo}{\cmdkl{\F}}
\knowledgenewcommand{\mapV}[1]{\cmdkl{\psi_{#1}}}
\knowledgenewcommand{\mapE}[1]{\cmdkl{\phi_{#1}}}

\section{Characterisation via attractor decompositions}\label{sec:strahler}

\subsection{Strahler number}

The Strahler number of a tree, given by the height of the largest full binary tree that appears as a minor, measures the arborescence of a tree. We generalise this notion.

\begin{figure}[!htb]
	\centering
	\begin{tikzcd}[sep=small]
		&&&&&& \bullet \\
		\\
		&& \bullet && \bullet && \bullet && \bullet && \bullet \\
		\\
		& \bullet & \bullet & \bullet & \bullet & \bullet & \bullet & \bullet & \bullet && \bullet && \bullet \\
		\\
		\bullet && \bullet & \bullet & \bullet & \bullet && \bullet & \bullet & \bullet & \bullet & \bullet & \bullet & \bullet
		\arrow[color={rgb,255:red,214;green,92;blue,92}, no head, from=1-7, to=3-3]
		\arrow[color={rgb,255:red,214;green,92;blue,92}, no head, from=1-7, to=3-5]
		\arrow[no head, from=1-7, to=3-7]
		\arrow[color={rgb,255:red,214;green,92;blue,92}, no head, from=1-7, to=3-9]
		\arrow[no head, from=1-7, to=3-11]
		\arrow[color={rgb,255:red,214;green,92;blue,92}, no head, from=3-3, to=5-2]
		\arrow[color={rgb,255:red,214;green,92;blue,92}, no head, from=3-3, to=5-3]
		\arrow[color={rgb,255:red,214;green,92;blue,92}, no head, from=3-3, to=5-4]
		\arrow[color={rgb,255:red,214;green,92;blue,92}, no head, from=3-5, to=5-5]
		\arrow[no head, from=3-7, to=5-6]
		\arrow[no head, from=3-7, to=5-7]
		\arrow[no head, from=3-9, to=5-8]
		\arrow[color={rgb,255:red,214;green,92;blue,92}, no head, from=3-9, to=5-9]
		\arrow[no head, from=3-11, to=5-11]
		\arrow[no head, from=3-11, to=5-13]
		\arrow[no head, from=5-2, to=7-1]
		\arrow[no head, from=5-3, to=7-3]
		\arrow[color={rgb,255:red,214;green,92;blue,92}, no head, from=5-5, to=7-4]
		\arrow[color={rgb,255:red,214;green,92;blue,92}, no head, from=5-5, to=7-5]
		\arrow[color={rgb,255:red,214;green,92;blue,92}, no head, from=5-5, to=7-6]
		\arrow[color={rgb,255:red,214;green,92;blue,92}, no head, from=5-9, to=7-8]
		\arrow[color={rgb,255:red,214;green,92;blue,92}, no head, from=5-9, to=7-9]
		\arrow[color={rgb,255:red,214;green,92;blue,92}, no head, from=5-9, to=7-10]
		\arrow[no head, from=5-11, to=7-11]
		\arrow[no head, from=5-11, to=7-12]
		\arrow[no head, from=5-13, to=7-13]
		\arrow[no head, from=5-13, to=7-14]
	\end{tikzcd}
	\caption{An "ordered tree" of depth 4, of 3-"Strahler number" 3, as exemplified by the red edges.}
	\label{fig:example-Strahler}
\end{figure}
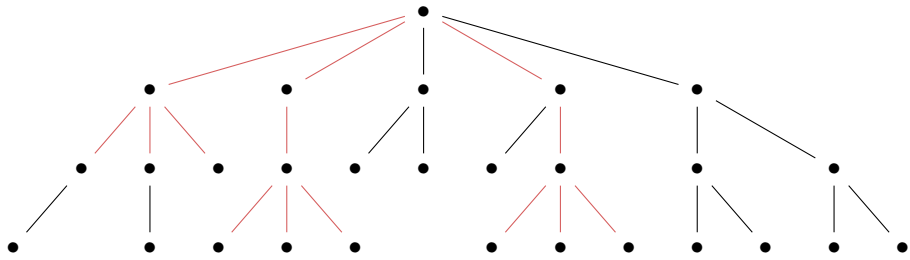

\AP Let $n\in \NN$. The ""$n$-Strahler number"" of $T$ a tree of finite depth, denoted $\intro*\Sn{n}T$, is defined by recurrence:
\begin{itemize}
	\item if $T = \langle \rangle$, $\Sn n T=1$.
	\item Else, $T = \langle (T_k)_{k\in K}\rangle$. We consider $m := \max\{\Sn n {T_k}|k \in K\}$. If there are at least $n+1$ $T_k$'s of "$n$-Strahler number" $m$, $\Sn n T = m+1$. Else, $\Sn n T = m$.
\end{itemize}
The "$n$-Strahler number" of $T$ is at most its "depth". Having a "$n$-Strahler number" $k$ is equivalent to having a complete $n$-ary tree of detph $k$ as a minor, for the operations of child deletion and replacing a node by one of its "children".  \cref{fig:example-Strahler} gives an example.

\AP We say that a "parity game" $G$ has ""$n$-Strahler number@@game-Strahler"" $j$ if there exists a strategy $\sigma_G$, winning for Eve, such that the resulting "parity graph" admits an "attractor decomposition" of tree-shape whith "$n$-Strahler number" $j$.

In the next two sections we prove each direction of the following theorem, using~\cref{thm:feasible-register} for the upward implication.

\thmparityattractors*

\begin{remark}Note that this theorem, based on a range $[1,2j]$, is less precise than~\cref{thm:feasible-register}, which handles all ranges $J$. This is because the parity of the minimal and maximal priorities are not reflected in the tree-shape of the attractor decomposition. For example, if there is a uniform bound on the lengths of paths in attractors, then there is no need for a minimal odd priority. The maximal even priority on the other hand is not required if there are no edges that go from $A_i$ to $A_j$ with $j>i$.  While the extremal parities are hard to characterise from the attractor decompositions, they are neatly captured by the transduction game.
	\end{remark}

\subsubsection{From feasibility to attractor decompositions}

Let $A$ be a "guidable automaton" of index $I$. If $\Lang(A)$ is $[1,2j]$ feasible by some automaton $B$, by~\cref{lem:guidable-n-bound}, there exists $n\in \NN$ and a "run" $\rho_B$ "guiding" $A$ such that the resulting "run" $\rho_A$ is "$n$-bound" by $\rho_B$. From this, we exhibit an "attractor decomposition" of $G$ of "$n$-Strahler number" $j$. More precisely these "runs" over $\AGame{A}{t}$ and $\AGame{B}{t}$ are considered as an "$I$-tree" and a "$[1,2j]$-tree", respectively. We will use these two "trees" in order to exhibit an "attractor decomposition" of $\AGame{A}{t}$ of "$n$-Strahler number" $j$.

\begin{restatable}{proposition}{propRegToStrahler}\label{lem:Reg-to-Strahler}
Given a tree-shaped graph $G=(V,E)$, "finitely-branching" and without terminal vertices,
two indices $I=[0,2i]$ and  $J=[1,2j]$ and
labellings $\rho_I:E\rightarrow I$ and $\sigma_J: E\rightarrow J$
such that $(G,\rho_I)$ and $(G,\sigma_J)$ are "even" "parity graphs",
if $\rho_I$ is "$n$-bound" by $\sigma_J$, 
then $(G,\rho_I)$ admits an "attractor decomposition" of "$n$-Strahler number" at most $j$.
\end{restatable}

\begin{proof}[Proof sketch]
In this proof, we begin with an "attractor decomposition" $(H,A_0, (S_k,A_k,D_k)_{k<\kappa})$ of $G_I$, the graph $G$ labelled by the run $\rho_I$. We then use $\sigma_J$ to refine this decomposition.

Within each $S_k$, we identify the vertices $S^*_k$ such that the path leading up to them has seen $2j$ since entering $A_k$. We then partition and order the sets $S^*_k$ into sets $\Theta_m$ such that a path that goes from one such set to another must see $2j$ in its $\sigma_J$ labelling and $2i-1$ in its $\rho_I$ labelling. The "$n$-boundedness" condition guarantees that there are no more than $n$ of these sets. These sets, with priorities in $\rho_I$ and $\sigma_J$ bounded by $2i-2$ and $2j$ respectively, have "attractor decompositions" of "$n$-Strahler number" up to $j$.

The remaining vertices of $S_k$ form subgames in which $2j$ does not occur, so they can be decomposed by an "attractor decomposition" following $\sigma_J$ ("even") into subgames in which priorities are dominated by $i-2$ and $j-2$: these admit "attractor decompositions" of "$n$-Strahler number" up to $j-1$, by induction hypothesis.

Then, assembled into the appropriate order, these up to $n$ "attractor decompositions" of "Strahler number" up to $j$ and arbitrarily many "attractor decompositions" of "$n$-Strahler number" up to $j-1$ are used to display the "attractor decomposition" of "$n$-Strahler number" at most $j$.

The details of this proof, 
\ificalp
in the full version,
\else
in~\cref{app:Reg-to-Strahler},
\fi
get quite technical, as it handles two different types of sub-decompositions that must be interleaved in the right order, with the appropriate attractors computed in between, while checking that all of the built sets satisfy the requirements to be in an attractor decomposition.
\end{proof}

\subsubsection{From attractor decomposition to feasibility}

For the backward direction of~\cref{cl:parity-attractors}, we show that if Eve has a winning strategy in a game with a corresponding attractor decomposition of "$n$-Strahler number" $h$, then she can win the corresponding "priority transduction game" with $h$ registers and counters going up to $n$. Then, using~\cref{thm:feasible-register}, we obtain the required implication.

\begin{restatable}{proposition}{strahlerToReg}\label{lem:Strahler-to-Reg}
Given a game $G$ and $n\in \NN\setminus\{0\}$, if $G$ has "$n$-Strahler number@@game-Strahler" $h$, then Eve wins $\Reg{[1,2h]}{n+1}(G)$.
\end{restatable}

\begin{proof}[Proof sketch]
Given an "attractor decomposition" of $G$ of $n$-Strahler number $h$, we build a winning strategy for Eve in $\Reg{[1,2h]}{n+1}(G)$.

The idea of her strategy is that when the underlying "parity game" takes an edge $(q,q')$, Eve identifies the smallest sub-"attractor decomposition" that contains both $q$ and $q'$. For technical reasons, if the priority of the move is odd and smaller than the maximal odd priority in the sub-"attractor decomposition", then she picks for her choice of priority in $I$ the said maximal odd priority. Otherwise, she uses the actual priority of the move.

 If the edge advances to the left in the "attractor decomposition", Eve picks the smallest register $r_0$, if it advances to the right (and hence is labelled with a relatively large even priority), she picks the register corresponding to the "$n$-Strahler number" of the sub-"attractor decomposition". If it stays within the same attractor, she picks $r_0$ or $r_1$ depending on the priority of the move. 

The technical part of the proof, detailed 
\ificalp
in the full version,
\else
in~\cref{app:Strahler-to-Reg},
\fi
then consists of checking that this strategy is indeed winning. The main idea is that a play will eventually stay in some minimal sub-"attractor decomposition", where it will see a maximal even priority from $I$ infinitely often. Then, Eve's strategy ensures that the maximal register $r_j$ used infinitely often corresponds to the $n$-Strahler-number of this decomposition. Since there are at most $n$ children of the same $n$-Strahler number, the counters $c_{i,j}$ are only incremented up to $n$ times before being reset by the occurence of a higher even priority, thus avoiding seeing a large odd output infinitely infinitely often. \qedhere

\end{proof}

\begin{remark}
	 Eve also has a winning strategy in $\Reg{[1,2h]}{n}(G)$, but the proof is more elaborate, as we need to do a case analysis of the behaviour of the last counter incrementation.
\end{remark}

If  Eve has such an attractor decomposition over all the games $\AGame{A}{t}$ for $t\in \Lang(A)$, the corresponding $n$ is a uniform bound such that Eve wins all the $\Reg{[1,2j]}{n}$. From this, we conclude the proof of~\cref{cl:parity-attractors} using the upwards direction of~\cref{thm:feasible-register}.

\section{Characterisation via universal trees}\label{sec:universal}

We now show that the previous characterisations of "$J$-feasibility" of a "guidable automaton" $\A$ can be reformulated in terms of the existence of a "universal tree" for $\A$. Note that in this section we use both "trees", which are binary, infinite and inputs to automata, and "ordered trees", which are of potentially infinite branching but finite height and describe "attractor decompositions".

\AP We say that an "ordered tree" is ""universal for an automaton $\A$"" if it is "universal" for some set of "ordered trees" $T$ such that for all "regular trees" $t\in L(\A)$, Eve has a "strategy" in $\AGame{\A}{t}$ with an "attractor decomposition" of tree-shape in $T$.

\thmuniversaltrees*

To prove this theorem, we show that that for fixed $n,j,d\in \NN^+$, there is an infinite "ordered tree" $\U$ of "$n$-Strahler number" $j$ and "depth" $d$ that is "universal" for the set of finite "ordered trees" of $n$-Strahler number at most $j$ and "depth" at most $d$. Over regular "trees", because of the positionality of "parity games", Eve's strategies can be chosen to be regular, which implies that their "attractor decompositions" can be finite, making $U$ universal for guidable automata recognising a $[1,2j]$-feasible language. For the other direction, we recall that if two tree automata are equivalent over "regular trees", they are equivalent over all "trees"~\cite{RabinRegular}.\\

\AP Let $n,k,d\in \NN^+$. We define recursively the ""universal tree"" $\intro*\Ut{n,k,d}$
of "$n$-Strahler number" $k$ and "depth" $d$ as follows, where $\omega(T)$ denotes the repetition of $\omega$ times the "ordered tree" $T$:

\begin{itemize}
	\item $\Ut{n,1,1} := \langle \rangle$
	\item When $d<k$, $\Ut{n,k,d}$ is undefined.
	\item Else, $d\geq k$, and by denoting $U := \Ut{n,k-1,d-1}$, we have \\$U_{n,k,d} := \langle \omega(U), \Ut{n,k,d-1}, \omega(U), \dots, \Ut{n,k,d-1}, \omega(U)\rangle$, with $n$ repetitions of $\Ut{n,k,d-1}$ (or, if it is not defined, no such repetition). Similarly, if $U = \Ut{\alpha,n,k-1,d-1}$ is undefined due to $k$ being equal to $0$, these children are omitted.
\end{itemize}

\begin{figure}[!htb]
	\includegraphics[width=\textwidth]{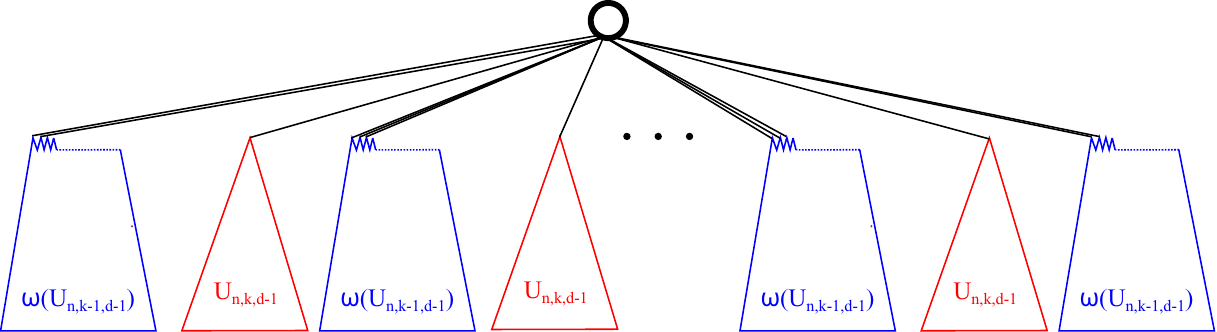}
	\caption{The recursion step in the construction of $\U_{\alpha,n,k,d}$.}
	\label{fig:univ-strahler}
\end{figure}

An example of such a construction can be found in figure \ref{fig:univ-strahler}.
Observe that $\Ut{n,k,d}$ has "width" greater than $\omega$ as soon as $2\leq k\leq d$.
Furthermore, $\forall n,k,d$, $\Ut{\alpha,n,k,d}$ has "depth" $d$, and we establish that it also has "$n$-Strahler number" exactly $k$:
\begin{lemma}
	For all $n,k,d\in \NN\setminus\{0\}$, if $k\leq d$, then $\Ut{n,k,d}$ is defined, and $\Sn{n} {\Ut{n,k,d}} = k$.
\end{lemma}
\begin{proof}
	We proceed by induction on $(d,k)$ ordered by the sum $d+k$.
	\begin{itemize}
		\item 	If $(d,k) = (1,1)$, the result is immediate.
		\item Else, by induction $\Ut{n,k-1,d-1}$ is defined and has "$n$-Strahler number" $k-1$, and if defined $\Ut{n,k,d-1}$ has "$n$-Strahler number" $k$. We observe that $\Sn n {\Ut{n,k,d}} = k$, as it has at most $n$ "children" of "$n$-Strahler number" $k$, and more than $n$ "children" of "$n$-Strahler number" $k-1$.\qedhere
	\end{itemize}
\end{proof}

We can now prove its "universality":

\begin{lemma}\label{lem:Ut-universal}
Let $\T$ be a set of finite "ordered trees", all of "depth" bounded by $d$ and "$n$-Strahler number" at most $k$. Then $\Ut{n,\min(k,d),d}$ is "universal for $\T$".
\end{lemma}
\begin{proof}
We reason by recurrence on $(d,k)$ ordered by the sum $d+k$.\\
If $k = d = 1$, then $\forall T \in \T, T = \langle\rangle$, and is trivially "isomorphically embedded" in $\Ut{n, 1,1}$.\\
Else, let $(d,k) \neq (1,1)$, and we suppose by recurrence that $\forall (d',k')$ such that $d'+k' < d+k$, the proposition holds. If $d<k$, then there is no "ordered tree" of "depth" $d$ and "$n$-Strahler number" $k$: they all are of $n$-Strahler number at most $d$. Then, by recurrence, as $2d < d+k$, $\Ut{n,d,d}$ is "universal for $\T$".

Else, $\Ut{n,k,d} = \langle \omega(\Ut{n,k-1,d-1}), \Ut{n,k,d-1}, \omega(\Ut{n,k-1,d-1}), \dots, \Ut{n,k,d-1}, \omega(\Ut{n,k-1,d-1})\rangle$, with $n$ repetitions of $\Ut{\alpha,n,k,d-1}$ if it is defined (else zero such repetition). Let $T\in \T$. By definition, if $T = \langle\rangle$, it is "isomorphically embedded" in all "ordered trees". Else, for $T = \langle (T_i)_{i\in I}\rangle$, then $\Sn{n}{T}\leq k$, and notably it admits at most $n$ $T_i$'s of "$n$-Strahler number" $k$. We denote the corresponding indices $(i_1, \dots, i_m)$ with $m\leq n$. Denoting $(j_1, \dots, j_n)$ the (ordinal) indices of the $\Ut{n,k,d-1}$'s in  $\Ut{n,k,d}$, we define $\psi: i_l \mapsto j_l$, for all $l\leq m$. By recurrence, we have that $\forall l \leq m, T_{i_l}$ is "isomorphically embedded" in $U_{\alpha,n,k,d-1}$ (as $d+(k-1) < d+k$). All the other $T_i$'s are such that $\Sn n {T_i} \leq k-1$, and are thus "isomorphically embedded" in $\Ut{\alpha,n,k-1,d-1}$ by recurrence (if $k-1 \neq 0$, that is. If $k-1=0$, these $T_i$ do not exist, as they would be of "$n$-Strahler number" $0$). Then, defining $i_0 = -1$ and $i_m = \omega$ as for all $l \in [0,m]$ there is only a finite number of such "ordered trees" $T_i$ between the indices $i_l$ and $i_{l+1}$, we can easily map in order $(T_i)_{i\in [i_l, i_{l+1})}$ in the corresponding $\omega(\Ut{n,k,d-1})$ with a map $\phi_l$. We finally observe that the function obtained by combining $\psi$ and the different $\phi_l$ is indeed injective, increasing, and that it maps $T_i$'s to "ordered trees" in which they are "isomorphically embedded", and thus describes an "isomorphic embedding" of $T$ in $\Ut{n,k,d}$.
\end{proof}

\begin{remark}\label{rem:regular-generalises}
As established by Rabin~\cite[Theorem 20]{RabinRegular}, a non-empty tree automaton accepts a regular tree, therefore, if two "automata" are equivalent in $\RgS_\Sigma$, they are equivalent over all "trees". This notably implies that, for $A$ an automaton and $J$ an index, $\Lang(A)$ is $J$-"feasible" over $\RgS_\Sigma$ if and only if it $J$-"feasible" over all "trees".
\end{remark}

\begin{lemma}
	Let $A$ a "guidable" "$I$-automaton", let $t$ a "regular tree" in $\Lang(A)$. If $\Lang(A)$ is $[1,2j^*]$-feasible, then there exists a finite "attractor decomposition" of $\AGame{A}{t}$ of "$n$-Strahler number" at most $j^*$.
\end{lemma}
\begin{proof}
	As $\Lang(A)$ is $[1,2j^*]$-feasible, there exists a "$[1,2j^*]$-automaton" $B$ that recognizes $\Lang(A)$. As $t$ is a "regular tree", using the positional determinacy of parity games, we can exhibit the existence of an "accepting run" $\rho_B$  of $B$ such that $\rho_B$ is a "regular tree". From this, we obtain that the "run" $\rho_A$ of $A$ "guided by" $\rho_B$ is also "regular". There thus exists $G_{\rho_A}$ finite graph whose unfolding is $\rho_A$, similarly there exists $G_{\rho_B}$ of unfolding $\rho_B$. Neither of  them has any terminal vertices, else it would imply the existence of terminal vertices in $\rho_A$ or $\rho_B$.
	
	We consider the graph $G = G_{\rho_A} \times G_{\rho_B}$, still finite and without terminal vertices. We then define $G'$, consisting of $G$ with some memory $M$: for each $2i-1\in I, 2j \in [1,2j^*]$, it stores whether a $2j$ was seen in its $\rho_B$ component since the last $2i-1$ in its $\rho_A$ component, and, conversely, whether $2i-1$ has been seen in the $\rho_A$ component since the last $2j$ in the $\rho_B$ component. We denote $L_A$ and $L_B$ the labelling fonctions of $G'$ in $I$ and $[1,2j^*]$, respectively. We observe that unfolding $G'$ on $L_A$ is still induced by the "run" $\rho_A$, and similarly with $L_B$ and $\rho_B$. Then, by \cref{lem:guidable-n-bound}, $L_A$ is $n$-bound by $L_B$.
	
	We can then apply a variant of \cref{lem:Reg-to-Strahler} on $L_A$ and $L_B$ with underlying graph $G'$. The graph $G'$ being finite, it does not satisfy the tree-shaped condition; however this hypothesis is used only twice in the initial proof, to find vertices to which every path has seen a $2j$ since entering the current $A_k$ and vertices to which every path has seen $2i-1$ since the last $2j$ (through a property we call "tightness"). As the memory $M$ stores exactly this information, we can instead define $S^*_k$ as the vertices in $S_k$ that saw a $2j$ since the last $2i-1$, and increases in the "star-rank" now only take place if the memory states that a $2i-1$ was seen since the last sighting of a $2j$ (intuitively, a state belongs to some $\Theta_m$ if and only if its memory states that a $2j$ was seen last). The remainder of the proof is identical, and it still builds an "attractor decomposition" of $G'$ of "$n$-Strahler number" $j^*$, which is finite since $G'$ is finite.\qedhere
	
\end{proof}

We finally obtain the direct implication of~\cref{cl:universal-trees} from this lemma and \cref{lem:Ut-universal}.

For the converse direction, by~\cref{lem:Strahler-to-Reg}, Eve wins all the $\Reg{[1,2j]}{n+1}(\AGame{A}{t})$ for $t\in \Lang(A)\cap \RgS_\Sigma$. Conversely, for $t \in \Lang(A)^C \cap \RgS_\Sigma$, by~\cref{lem:reg-rejects-rejecting}, Eve looses. Therefore, by \cref{lem:automata-composition} (restricting ourselves to the "regular trees"), we can construct $B$ a $[1,2j]$-automaton recognizing $\Lang(A)\cap\RgS_\Sigma$ over the "regular trees". Therefore $\Lang(A)$ is $[1,2j]$-feasible over the "regular trees", and by~ \cref{rem:regular-generalises} is thus $[1,2j]$-feasible.

\section{Conclusion}
We have given three closely related new characterisations of the "$J$-feasibility" of "$\omega$-regular tree languages": one via the "transduction game", one via "attractor decompositions" and one via "universal trees". While we do not solve the decidability of the index problem, our work brings to light the deep relationships between the tools we are used to manipulate in the context of solving "parity games", such as "attractor decompositions", "universal trees" and Lehtinen's register game, and the "$J$-feasibility" of a language.  In particular, the "$n$-Strahler number" turns out to have great explanatory power by relating the "transduction game", the structure of "attractor decompositions" and the "index" of a language. 

The B\"uchi case, which is at the frontier of the state of the art, is particularly appealing because of its simplicity: the language of a "guidable automaton" $\A$ is B\"uchi "feasible" if and only if there is a finite bound $n$ such that Eve can win in the acceptance games with strategies with "attractor decompositions" of width at most $n$, or, equivalently, if $\A$ admits a \textit{finite} "universal tree". We hope that these insights will help unlock the next steps in tackling this long-standing open problem.

\bibliography{biblio}

\appendix

\section{Appendix}
% !TEX root =  main.tex

\subsection{Proofs from~\cref{sec:register-game}}

\automataComposition*
\begin{proof}
	Let $C$ an automaton describing the "transduction game" $\Reg{J}{n}$, that is, accepting "trees" $\rho$ such that Eve wins $\Reg{J}{n}(\rho)$. 
	That is, $C$ has for states the set $Q_C$ the set of the different configurations of $\Reg{J}{n}$, which are in finite number, and on an input tree $\rho$, $C$ behaves like $\Reg{J}{n}(\rho)$. Formally, for $i\in I$ (the label of the current position in $\rho$) and a configuration $c$, let $\C$ the sets configurations that can be reached by Eve in one step in $\Reg{J}{n}$ (from the point where Adam made his move, bringing us to the position of label $i$). Then $\delta_C(c,i) = \C$, each transition being labelled with the corresponding output $w \in J$. We observe that on each "branch" $b \in \{0,1\}^\omega$, the different "runs" in $C$ correspond to the different plays of $\Reg{J}{n}(\rho)$ where Adam successively chose the directions of $b$.
	Let us design $B$ as the composition of $A$ and $C$. That is, $B$ takes for input some $\Sigma$-"tree" $t$, on which $A$ admits a "run" $\rho$. Then $C$, taking $\rho$ as input, accepts if Eve wins $\Reg{J}{n}(\rho)$. We observe that $B$ has states $(q,c) \in Q_A \times Q_C$, hence, as $Q_C$ is finite, this composition indeed forms an "automaton" of "index" $J$. Let us show that it recognizes exactly $\Lang(A)$.\\
	If $t\notin \Lang(A)$, all the "runs" of $A$ on $t$ are "rejecting", hence by \cref{lem:reg-rejects-rejecting}, the output of $\Reg{J}{n}$ is rejecting on any "run" of $A$ on $t$. Else, $t\in \Lang(A)$, and by hypothesis, Eve wins $\Reg{J}{n}(\AGame{A}{t})$. Therefore, for $\rho_t$ the "run" of $A$ that she uses to win $\AGame{A}{t}$, Eve can follow her strategy in $\Reg{J}{n}(\AGame{A}{t})$ to resolve the non-determinism of $C$. Therefore $t \in \Lang(B)$. Hence, $\Lang(B) = \Lang(A)$, and $\Lang(A)$ is $J$-"feasible".
\end{proof}

\regRejectsRejecting*

\begin{proof}
	As $G$ is not "even", there exists an infinite path in $G$ dominated by some odd $\hat{i}$.
	In $\Reg{J}{n}(G)$, Adam will simply follow this path. Therefore, the sequence $L(e)$ is dominated by $\hat{i}$, and thus the sequence of $i\in I$ chosen by Eve is dominated by some odd $i^*\geq \hat{i}$. This $i^*$ is therefore maximal among all "priorities" seen after the step $n_0$, for some $n_0 \in \NN$.\\
	Let $(j_n)_{n\in \NN}$ the infinite sequence of "registers" chosen by Eve, dominated by some $j^*$, therefore maximal among all "register" indices chosen after some $n_1 \geq n_0 \in \NN$. Let us look at events past the $n_1$-th step of the game.\\
	If $j^* = 0$, the game only outputs $\min(J) = 1$, odd, and Adam indeed wins. We can thus restrict ourselves to the case were $j^* \neq 0$.
	Infinitely often, $\rho(p) = i^*$, hence, for $j_0$ picked at the corresponding step, $\forall j' \geq j_0, r_{j'}\geq i^*$. It is notably the case for $r_{j^*}$, by maximality of $j^*$. Therefore, infinitely often, as a $j^*$ will recur and that $r_{j^*}$ cannot be reduced before then (as the value of a "registers" is non-decreasing until it is chosen), we will see $r_{j^*} = i^*$. At each such moment, we are in one of the two latter case of the case disjunction : either we output $2j^*+1$, either we increment $c_{i^{*},j^*}$. Along the "branch" $b$ we no longuer see a value superior to $i^*$, nor pick a "register" superior or to $j^*$. Hence, the "counter" $c_{i^*,j^*}$ is never reset : we thus output infinitely often a $2j^{*}+1$ (or even immediately lose if $2j^* = \max(J)$).
	As we no longer pick any "register" $> j^*$, we easily see that we never output any "priority" $>2j^*+1$, which concludes as to the fact that the output sequence is "rejecting".
\end{proof}

\pumping*

\begin{proof}
Given an infinite tree $t$, we define the tree $t^*$ starting from $t$, where, for $t_u$ the subtree of $t$ at position $u$, for all $n \in \NN$, $t_u$ replaces recursively the subtrees at positions $u \concat v^n$.\\
We use the same construction to define the "runs" $\rho_A^*$ and $\rho_B^*$. They are legitimate "runs" on $t^*$, as $\rho_B(u) = \rho_B(u\concat v)$, therefore at each repetition of $t_u$, $A$ is in the same state $q_u$ and can thus choose the same transition. The same applies to $\rho_B$.

Let $b^*$ the unique "branch" going through all the repetitions of $t_u$ in $\rho_B^*$. We observe  that this corresponds to the "branch" $u \concat v^\omega$. Therefore, past the position $u$, $b^*$ infinitely repeats the segment from $u$ to $u\concat v$ of the subtree $\rho_B(u)$. This segment is dominated by an even priority $p$ by lemma hypothesis. Therefore $b^*$ is dominated by $p$ even, and is thus "accepting@accepting run".

We observe that on  any other "branch" $b$ of $\rho_B^*$ (that is, $b$ is of the shape $u\concat v^k \concat w$ with $w \neq v^\omega$), the suffix of "path@@tree-p" along $b$ in $\rho_B^*$ is a clone of the "path@@tree-p" along $u\concat w$ in $\rho_B$, and is thus "accepting". We thus obtain that $\rho_B^*$ is "accepting@accepting run".

Let us denote $g$ the "guiding function" $g : Q_A \times \Delta_B \to \Delta_A$, $(q_i)_{i < |v|}$ the states taken by $A$ between $u$ and $u \concat v$, and $(\delta_i)_{i < |v|}$ the transitions taken in $B$ between $u$ and $u \concat v$. We have, as $\rho_A$ is "guided by" $\rho_B$, that on this (repeated) segment, $\rho_A$ takes the transitions $(g(q_i, \delta_i))_{i< |v|}$.
Hence, as $q_0$ is repeated at position $u\concat v$, and in $\rho_B'$ the same transitions are repeated along this "branch", we obtain that the "run" $\rho_A^*$ is a run guided by $\rho_B^*$. Notably, as $\rho_B^*$ is "accepting@accepting run", so is $\rho_A^*$: therefore, on the "branch" $b^*$ along $u \concat v^\omega$, $\rho_A^*$ is dominated by an even priority, hence an even priority dominates $(q_i)_{i<|v|}$.
\end{proof}

\guidableNBound*

\begin{proof}	
	If by contradiction there exists a path $\pi$ in $t$, segmented into consecutive paths $\pi_0,\pi_1,\dots \pi_n$, such that for some even $j$ and some odd $i$, the maximal priority on the $L_I$- and $L_J$-labels of each $\pi_m, m\in [0,n]$ are $i$ and $j$, respectively. For a position $p\in \{0,1\}^*$, we denote $q(p)$ the couples of states in $(Q_A,Q_B)$ in which the "runs" $\rho_A$ and $\rho_B$ respectively are. Looking at the starting points $(p_i)_{i\leq n}$ of the paths $(\pi_i)_{i\leq n}$, there are thus at least two different $i,j \in [0,n]$ such that $q(p_i) = q(p_j)$. Between these two points, by construction of $\pi$, $\rho_B$ is dominated by an even $j$. Then, by \cref{cl:pump}, this segment is dominated by some even $i'$, contradiction with the fact that it would be dominated by $i$ odd. Therefore, there does not exist such a path $\pi$ in $t$ – and thus, there does not exist such a path in the labellings induced by $\rho_A$ and $\rho_B$.
\end{proof}

\subsection{Proofs from~\cref{sec:strahler}}
\subsubsection*{Proof of \cref{lem:Reg-to-Strahler}}\label{app:Reg-to-Strahler}

We begin with three technical lemmas: \cref{lem:attr-union} on the "attractor" of a union of disjoint sets, and \cref{lem:join-attr-decomp} that builds an "attractor decomposition" once all the subparts $S_k$ have been identified. The last one exhibits an "attractor decomposition" with more structure than usual, in our specific case where the underlying graph is a "tree".

\begin{lemma}\label{lem:attr-union}
	Let $\kappa$ be an ordinal, let $G=(V,E,L)$ a "finitely-branching" "parity graph", and $(S_k)_{k<\kappa}$, $\kappa$ disjoint subsets of $V^*$. Then, Then, by defining iteratively, for $k \leq \kappa$, $V_k:=V\setminus \bigcup_{j< k}A_j$ and $A_k := \attr(S_k,G\restrict V_k)$, we have that $\attr(\bigsqcup_{k<\kappa}S_k,G) = \bigsqcup_{k<\kappa} A_k$.
\end{lemma}
\begin{proof}
	We first observe that $\bigcup_{k<\kappa} A_k$ is always a disjoint union. Indeed, let $v\in \bigcup_{k<\kappa} A_k$, let $k_0$ the smallest $k$ such that $v\in A_k$. Then $\forall k' \geq k_0$, $A_{k_0} \cap V_{k'} = \emptyset$, hence $v \notin V_{k'}$ – therefore $v\notin A_{k'}$. Each $v \in \bigsqcup_{k<\kappa} A_k$ therefore belongs to a single element of the union.\\
	The converse inclusion is immediate : for $v\in \bigsqcup_{k<\kappa} A_k$, it notably belongs to a single $A_{k_0} = \attr(S_{k_0},G\restrict V_{k_0})$, and thus belongs to $\attr(\bigsqcup_{k<\kappa}S_k,G)$ as all its exiting paths eventually pass by an $S_k$ with $k\leq k_0$.\\
	For the direct inclusion, we proceed by transfinite induction on $\kappa$, for any parity graph $(G,E,L)$.
	\begin{itemize}
		\item If $\kappa = 1$, there is a single $S_0$ and the result is immediate.
		\item If $\kappa = n+1$ with $1\leq n$, supposing the result true up to rank $n$:
		let $v\in \attr(\bigsqcup_{k<n+1}S_k,G)$. If it has no successors in $\bigsqcup_{k<n}S_k$, then necessarily all its exiting paths eventually pass by $S_n$. Else, whether it may have paths ending in $S_n$ or not, it belongs to $\attr(\bigsqcup_{k<n}S_k,G\restrict (V \setminus A))$. Thus $\attr(\bigsqcup_{k<n+1}S_k,G) \subseteq A \sqcup \attr(\bigsqcup_{k<n}S_k,G\restrict (V \setminus A)) = A \sqcup \bigsqcup_{k<n} A_k$ by induction hypothesis.
		\item If $\kappa$ is the limit ordinal:\\
		Let $v \in \attr(\bigsqcup_{k<\kappa}S_k,G)$. We look at a set $\S \subseteq \{S_i|i<\kappa\}$ such that $v \in \attr(\bigcup_{S\in\S}S,G)$, taken minimal for the inclusion. Let $\kappa_v := \sup_{S_i\in \S}(i)$. If $\kappa_v < \kappa$, by induction hypothesis, $v \in \bigsqcup_{k<\kappa_v} \attr(S_k,G\restrict V_k)$. We observe that $v \notin \attr(S_{k'},G\restrict V_{k'})$ for $\kappa_v \leq k'$, as then $v \in V_{k'}$. We now show need to show that $\kappa_v$ cannot be a limit ordinal (and notably, cannot be equal to $\kappa$). As there are infinitely many (disjoint) $S_k$, when we look at the subgraph $G_v$ induced by the successors of $v$ in $G$, where all the $S_k$ are replaced by sink vertices. $G_v$ is "finitely branching", infinite and connected. Therefore, by K\"onig's lemma, there exists an infinite path from $v$ in $G_v$ without repeated vertices, therefore the corresponding path in $G$ is an infinite path that never encounters any $S_k$: contradiction as then $v\notin \attr(\S,G)$.
	\end{itemize}
\end{proof}

\begin{lemma}\label{lem:join-attr-decomp}
	Let $G = (V,E,L)$ a "finitely-branching" "parity game" of maximal even parity $h$. Let $H$ be the set of its transitions labelled by $h$, and $A_0 := \attr(H,G)$. and $(S_k)_{1\leq k\leq\kappa}$ a family of disjoint subsets of $V$. If $(S_k)_{1\leq k \leq \kappa}$ is such that
	\begin{itemize}
		\item $\forall 1 \leq k \leq \kappa$, $S_k$ is closed under successor in $(G\setminus H)\setminus \attr(\bigcup_{k'<k} S_{k'}, G\setminus H)$.
		\item $\forall 1\leq k \leq \kappa, (G\setminus H)\restrict S_i$ is a subgame containing priorities up to $h-2$, with an "attractor decomposition" $D_i$ of level $h-2$
		\item $\attr(\bigcup_{0\leq k\leq\kappa}S_i,G\setminus H) = V$.
	\end{itemize}
	Then, by defining iteratively, for $1\leq k \leq \kappa$, $V_k:=V\setminus \bigcup_{j< k}A_j$ and $A_k := \attr(S_k,(G\setminus H)\restrict V_k)$, $(H,A_0, (S_k,A_k,D_k)_{1\leq k \leq \kappa})$ is an "attractor decomposition" of $G$ (up to neglecting the empty $S_k$, of empty attractors).
\end{lemma}
\begin{proof}
	We observe that
	\begin{align*}
		\forall 1\leq k \leq \kappa, V_k &= V\setminus \bigcup_{j< k}A_j\\
		&= V\setminus \bigsqcup_{j< k}A_j\\
		&= V \setminus \attr(\bigsqcup_{j<k}S_k,G\setminus H),
	\end{align*} by \cref{lem:attr-union}, therefore $S_k$ is closed under successor in $(G\setminus H)\restrict V_k$. We have directly from the second item that all its transitions are bounded by $h-2$. We observe that $A_k$ indeed corresponds to $\attr(S_k,(G\setminus H)\restrict V_k)$.\\
	We still need to prove that $V_{\kappa+1} = \emptyset$.
	We have that $\attr(\bigcup_{0\leq k\leq\kappa}S_k,G\setminus H) = V$, thus, because the $S_k$ are disjoint, we observe that $V = \bigsqcup_{0\leq k \leq \kappa} A_k$. Therefore, $V_{\kappa+1} = \emptyset$.
	All the other conditions required for this tuple to form an attractor decomposition of $G$ being already satisfied by hypothesis, we conclude.
\end{proof}

\AP We say that an "attractor decomposition" $(H,A_0,(S_i,A_i,D_i)_{0<i<\ell})$ is ""tight"" if, for all $i,j < \ell$ with $j<i$, all paths from $S_i$ to $S_j$ in $G\setminus H$ are dominated by $h-1$, and the $(D_i)_{(0<i<\ell)}$ are also "tight".

\begin{lemma}\label{lem:tightAD}
	If $G$ is an "even" "parity game" such that the underlying graph is a "tree", we can build a "tight" "attractor decomposition" $D=(H,A_0,(S_i,A_i,D_i)_{0<i<\ell})$ of $G$.
\end{lemma}
\begin{proof}
	We build $D$ following initially the construction of an "attractor decomposition" used in the proof of \Cref{cl:EvenIffDecomposition}, although with a divergence in the middle.
	
	Let $H$ be the set of edges of priority $h$ and $A_0$ the "attractor" of $H$. 
	Then, we define each $S_i$ and $A_i$ for $i>0$ inductively. First let $V_i=V\setminus \bigcup_{j< i} A_j$ for $i\in \NN$  or an ordinal.
	$V_i$ is either empty, or $G_i=G\restrict V_i$ is an "even" "parity graph" with maximal priority no larger than $h{-}1$. Let $S'_i$ consist of all positions of $G_i$ from where $h{-}1$ can not be reached. That is, $S'_i$ is "even" (being a subgraph of $G$) and only has edges of priority up to $h-2$. 
	If $V_i$ is non-empty, there must be such positions, since otherwise one could build a path which sees infinitely many $h{-}1$, contradicting that $G_i$ is an "even graph". 
	
	We then define 
	$$S':= \{v \mid \exists j < i, v\in S_j, \exists \pi \text{ "path@@tree-p" from } S'_i \text{ to } v \text{ in }(G\setminus H) \text{ dominated by at most } 2h-2\},$$
	and define $S_i := S'_i \cup S'$ (and thus we remove vertices from $S'$ in all the corresponding $S_j$). Note that this induces a modification of some $(A_j)_{j<i}$, and thus of the corresponding $V_{j+1}$. It does not, however, modify the set of vertices in $A_i := \attr(S_i, G_i)$ (as we only put add $S_i$ vertices reachable from the vertices initially in $S_i$, and $G$ is a "tree").
	We need to show that the ensuing $(S_j)_{j\leq i}$ still satisfy the desired properties. We immediately observe that they still only contain edges of priority at most $h-2$, and that this operation has not created any terminal vertex. Additionally, as $G$ is a tree, for $v\in S'$ initially in some $S_j$, it has no predecessor in $G_j\setminus S'$ for paths in $G\setminus H$ : all the predecessors by a path dominated by at most $h-2$ are in $S'$, and as $v\in S'$, all its predecessors for paths dominated by $h-1$ are also predecessors of some vertex in $S'_i$, and thus is not in some $S_{j'}$ with $j'<i$: removing $v$ from $S_j$ thus does not break the closure by successor of such a $S_{j'}$. Finally, as we only added to $S'_i$ vertices without successors in $G_i$, we immediately observe that $S_i$ is closed by successor in $G_i$.
	Let $(D_j)_{j\leq i}$ be "tight" "attractor decomposition" of level $h{-}2$ of $(S_j)_{j\leq i}$, which we can exhibit by recursion.
	
	Note that as each vertex in $G$ only has a finite number of predecessors, it can belong to such an $S'$ only a finite number of times, and the (potential) $S_i$ to which it belongs eventually stabilises.
	
	As $A_{i}$ is unchanged by this operation, we observe, as in the proof of \Cref{cl:EvenIffDecomposition}, that $V=\bigcup_{i< \ell}A_i$.
	Therefore, $D$ indeed corresponds to an "attractor decomposition" of $G$.
	
	We easily observe that by construction, all the transitions between the different $(S_i)_{i < \ell}$ are at least of priority $h-1$, which gives the "tightness" of this "attractor decomposition", as all the $D_i$ are "tight" by recursion.
\end{proof}

We can then proceed with~\cref{lem:Reg-to-Strahler}, in which we build the attractor decomposition.

\propRegToStrahler*

\begin{proof}
	
	This proof is quite elaborate, and we first give an overview of how it will proceed. The proof works by induction on the pair $(i,j)$. The base case is easy. For the induction step, we start from an attractor decomposition of $G$ along $I$ (which exists, as $\rho_I$ is "even"). We partition the obtained attractors into smaller subsets, each with their own "attractor decomposition". The goal is to exhibit at most $n$ subsets whose "attractor decomposition" is of "$n$-Strahler number" $j$, for which the induction hypothesis proves useful. We finally use \cref{lem:join-attr-decomp} to build the desired "attractor decomposition".
	
	For any subgraph $F$ of $G$, we write $F_I$ and $F_J$ for its $\rho_I$- and $\sigma_J$-labelled versions, respectively, noting that $\rho_I$ and $\sigma_J$ are also $n$-close labellings of all subgraphs of $G$.
	
	We proceed by induction on $(i,j)$ with lexicographical order. For the base case, the lemma is trivially true for $i=0$ and all $j$ since a $[0]$-labelled graph has an attractor decomposition of $n$-Strahler number $1$.
	For the induction step, we will use the statement for $(i-1,j)$ and $(i-1,j-1)$ to prove the statement for $(i,j)$.
	
	\textbf{{Initial attractor decomposition:}} By \Cref{lem:tightAD}, there exists a "tight" "attractor decomposition" $(H,A_0, (S_k,A_k,D_k)_{k<\kappa})$ of $G_I$ for $\kappa$ an ordinal. Similarly as in the definition, we set $V_k=V\setminus \bigcup_{l< k}A_l$. We denote $G^\dagger := (G\setminus H)\restrict V_1$.
	
	Let $k<\kappa$. We observe that $S_k$ has no terminal vertex $v$. Indeed, $G$ has no terminal vertex, hence if by contradiction all the transitions from $v$ were leaving $S_k$, as $S_k$ is closed under successors in $(G\setminus H) \restrict V_k$, they are all going to attractors of different $(S_{l})_{l <k}$, hence $v \in A_{l'}$ for $l'$ the maximal index among these attractors. As $l' < k$, we would have that $v \notin V_k$ and thus $v\notin S_k$, contradiction.
	
	\textbf{{Decomposing attractors by rank:}} In a given $S_k$, we denote $S^*_k$ as the set of vertices $v\in S_k$ such that there exists a non-empty path $\pi$ in $G^\dagger \cap A_k$, ending in $v$, that sees a $2j$ in $G^\dagger_J$. As $G^\dagger$ is a tree, this path is unique. That is, the vertices of $S^*_k$ consist of the vertices of $S_k$ such that we saw a $2j$ in $G^\dagger_J$ since entering in $A_k$. It obviously has no terminal vertex, for similar reasons as $S_k$. We denote its attractor in $S_k$ as $A^*_k := \attr(S^*_k, G\restrict S_k )$.
	
	\AP Let $k < \kappa, v \in S^*_k$. We denote its ""star-rank"" $\intro*\rks(v)$ to be $1$ if there is no $k'<k, v'\in S^*_{k'}$ such that there is a path from $v$ to $v'$ in $G^\dagger$, else, for $V'$ the set of such $v'$, $\rks(v) = 1+\sup_{v'\in V'}(\rks(v'))$. 
	
	We observe that there is no vertex $v_{n+1}$ of "star-rank" $n+1$ or greater, as else for the paths $\pi_n, \dots, \pi_1$ (and vertices $(v_l)_{1\leq l\leq n}$) exhibiting the successive increases in $\rks$, each such path $\pi_l$ is dominated by $2i-1$ in $G_I$ (as it goes from a $S^*_k$ to a $S^*_{k'}$ with $k' < k$ and the "attractor decomposition" is "tight"). The path $\pi_l$ is also dominated by $2j$ in $G_j$, as $v_l$ is in some $S^*_k$ and thus $\pi_l$ encounters a $2j$ in $(S_{k'})_J$ before reaching $v_l$. The path $\pi_n \pi_{n-1}\dots \pi{n}$ would then contradict the $n$-closeness property.
	
	We then define, for $1\leq m\leq n$, $\Theta_m := \bigsqcup_{k < \kappa}\{v \in S_k\mid \rks(v) = m\} = \{v\in V \mid \rks(v) = m\}$. We observe that for $m' < m$, there cannot be any path in $G^\dagger$ from $\Theta_{m'}$ to $\Theta_m$: else, the corresponding origin vertex in $\Theta_m$ whould have "star-rank" greater than $m$. 
	Similarly, for $m\in[1,n]$ and $v,v' \in \Theta_m$, if there is a path from $v$ to $v'$ in $G^\dagger$, then there exists $k<\kappa$ such that $v,v' \in S^*_k$. Else, such a path would be the witness that $v$'s "star-rank" should be greater than $m$. We thus deduce that this path admits no $2i-1$ in $G^\dagger_I$.
	As $\Theta_m$ corresponds to a subgame without terminal vertices (as the union of such subgames) and with edges priorities bounded by $2i-2$ in $\rho_I$ and by $2j$ in $\sigma_J$, by induction, it admits an "attractor decomposition" $D_m$ of "$n$-Strahler number" at most $j$.
	
	\textbf{{Decomposing remaining vertices by reachable rank:}} \AP For $v$ in some $S_{k}\setminus A^*_k$ we define its ""rank"" $\intro*\rk(v)$ as $0$ if there is no $k'\leq k, v'\in S^*_{k'}$ such that there is a path from $v$ to $v'$ in $G^\dagger$, else, for $V'$ the set of such $v'$, $\rk(v) = \sup_{v'\in V'}(\rks(v'))$. Once more, we observe that there is no vertex of "rank" $n+1$ or greater. We also observe that for $v\in S_{k}$, all the successors of $v$ in $S_{k}\setminus A^*_k$ have same "rank" as $v$. We denote $S_{k}^{(m)} := \{v\in S_{k}\setminus A^*_k|\rk(v)=m\}$. Note that this set can be empty. However, we still have that it has no terminal vertices, as any such terminal vertices that would appear by removing $S_k^*$ are in $A^*_k$.
	For reasons akin to the $\Theta_m$ case, we observe that there is no path in $G^\dagger$ from $S_{k}^{(m)}$ to any $\Theta_{m'}$ or $S_{k'}^{(m')}$ for $m < m'$.
	
	\textbf{Obtaining the remaining attractor decomposition of smaller level:} Let $k<\kappa, m\in [0,n]$. We consider $S_k^{(m)}$: it admits no $2j$-transitions by definition of $S^*_k$, hence as $G^\dagger$ is "even" on all its infinite paths, it admits an attractor decomposition in $J$ of the shape $(\emptyset, \emptyset, (S^{(m)}_{k,p}, A^{(m)}_{k,p}, D^{(m)}_{k,p})_{l<\kappa_k})$ where all the $S^{(m)}_{k,p}$ are subgames with $J$-labels bounded by $2j-2$. Therefore, as they are subgames of $S_k$, their $I$-labels are bounded by $2i-2$, and by induction hypothesis, we can thus suppose each $D^{(m)}_{k,p}$ to have "$n$-Strahler number" at most $j-1$. We have, by definition of "attractor decompositions", that there is no path in $G^\dagger$ from $S^{(m)}_{k,p}$ to any $S^{(m')}_{k',p'}$ with $(k,p)$ lexicographically smaller than $(k',p')$.
	
	\textbf{Building the desired attractor decomposition:} We now define the tuple $\D$, and will establish that it is indeed an "attractor decomposition" of $G$ of $n$-Strahler number at most $j$ by \cref{lem:join-attr-decomp}. We pose as the corresponding ordered sequence of disjoint subsets the sequence $$(\S_l)_{l<\alpha} := ((S^{(0)}_{k,p})_{p<\kappa_k})_{k<\kappa}, \Theta_1, ((S^{(1)}_{k,p})_{p<\kappa_k})_{k<\kappa}, \Theta_2, \dots, \Theta_n, ((S^{(n)}_{k,p})_{p<\kappa_k})_{k<\kappa}.$$ Their corresponding "attractor decompositions" are $$ (\D_l)_{l<\alpha} := ((D^{(0)}_{k,p})_{p<\kappa_k})_{k<\kappa}, D_1, ((D^{(1)}_{k,p})_{p<\kappa_k})_{k<\kappa}, D_2, \dots, D_n, ((D^{(n)}_{k,p})_{p<\kappa_k})_{k<\kappa}.$$
	We observe that they are all disjoint: the different $(S_k)_{k<\kappa}$ are disjoint by definition of an attractor decomposition, and similarly for the $(S_{k,p})_{p<\kappa_k}$. As the $S^*_k$ are disjoint from the $S_{k,p}$, it is still the case for the different $\Theta_m$. Finally, partitionning depending on $\rks$ or $\rk$ provides a disjoint union, which concludes.\\
	We can thus define iteratively, for $l < \alpha$, $\V_l:=V\setminus (A_0 \cup \bigcup_{l'< l}\A_{l'})$ and $\A_l := \attr(S_l,(G\setminus H)\restrict \V_l)$.\\
	As subgraphs of the $S_k$, we easily observe that all the $\S_i$ only admit $I$ transitions bounded by $2i-2$, and that for all $l<\alpha$, $\D_l$ is indeed an attractor decomposition for $\S_l$.\\
	We also observe that $(\S_l)_{l<\alpha}$ form a partition of the $(S_k)_{k<\kappa}$ (except for the vertices in some $A^*_k \setminus S^*_k$).
	We define $(V_k)_{k< \kappa}$ according to \cref{lem:attr-union} for the sequence of disjoint subsets $(S_k)_{k<\kappa}$. Then this lemma entails that 
	\begin{align*}
		\attr(\bigsqcup_{l<\alpha}\S_l, G^\dagger) &= \bigsqcup_{l<\alpha} \attr(\S_l,(G^\dagger)\restrict \V_l)\\
		&= \bigsqcup_{k<\kappa}(A^*_k\cup \attr(S^*_k, G^\dagger \restrict V_k)) \ \cup \bigsqcup_{k<\kappa}\attr(S_k\setminus A^*_k, G^\dagger \restrict V_k)\\
		&= \bigsqcup_{k<\kappa} \attr(S_k, G^\dagger \restrict V_k)\\
		&= V\setminus A_0.
	\end{align*}
	\textbf{Verifying the closeness by successor:}We still need to prove that all the $\S_l$ are closed by successor in $(G\setminus H)\setminus \attr(\bigcup_{l'<l} \S_{l'}, G\setminus H)$. We reason by case disjunction on $\S_l$.
	\begin{itemize}
		\item If $\S_l$ is of the shape $S^{(m)}_{k,p}$: by definition, $S_{k,p}$ is closed by successor in $(G\setminus H)\setminus \attr(\bigcup_{k'<k} S_{k'}\ \cup \bigcup_{p'<p}S^{(m)}_{k,p'}, G\setminus H)$. We still need to prove that it admits no vertex towards the $(\theta_{m'})_{m'>m}$, nor towards the $(S^{(m')}_{k',p'})_{m'>m}$. If either if these would be true, then for $v\in S^{(m)}_{k,p}$ at the origin of such a path, it would admit a successor of "star-rank" $m'>m$ (by transitivity in the second case), and thus $v$ would not be of "star-rank" $m$: contradiction.\\
		We finally observe that if there were a path from $\S_l$ towards a vertex $v \in \A_l \setminus \S_l$ in $(G \setminus H) \restrict \V_l$, as $\S_l \subseteq S_k$ (and admits no path in $(G \setminus H) \restrict \V_l$ towards $S^*_k$ nor the other $S^{m'}_{k,p'}$), it would imply a path from $S_k$ towards $v\notin S_k$. However, as $v\in \A_l$, there exists a path from $v$ towards $S_k$, hence $v \notin V_k$: contradiction with the fact that $S_k$ is closed by successor in $(G\setminus H)\restrict V_k$.
		\item If $\S_l$ is of the shape $\Theta_m$: if by contradiction it admitted a successor in a $\Theta_{m'}$ or $S^{(m')}_{k,p}$ with $m<m'$, for similar reasons, it would bring a contradiction as to the "star-rank" of some of its vertex. If by contradiction there exists in $G^\dagger$ a path from some $v'\in \Theta_m$ to some $v \in S^{(m)}_{k,p}$: then by definition of $S^*_k$, we observe that $v'\notin S^*_k$, and thus $\exists k' > k: v'\in S^*_{k'}$. As $v\in S^{(m)}_{k,p}$, it is of "rank" $m$, and thus admits a successor $v''$ of "star-rank" $m$, in some $S_{k''}$ with $ k'' \leq k$. Therefore, $v'$ has a successor $v''$ of "star-rank" $m$ in some $S_{k''}$ with $k'' < k'$: it is thus of "star-rank" at least $m+1$, contradiction. Therefore there does not exist such a successor $v$.\\
		We obtain for similar reasons as above that there does not exists in $(G \setminus H) \restrict \V_l$ a path from $\S_l$ towards a vertex $v \in \A_l \setminus \S_l$. Thus, $\S_l$ is indeed closed by successor in $(G\setminus H)\restrict \V_l$.
	\end{itemize}
	Then, by \cref{lem:join-attr-decomp}, $\D$ is indeed an "attractor decomposition" of $G$, and we observe that among the $\D_l$, it has at most $n$ of them are of "$n$-Strahler number" $j$. Hence $\D$ is indeed an "attractor decomposition" of "$n$-Strahler number" $j$.
\end{proof}

\subsubsection{Proof from~\cref{sec:universal}}\label{app:Strahler-to-Reg}

\strahlerToReg*

\begin{proof}
	Let $\sigma_G$ be the corresponding "winning strategy" of Eve in $G$, with "attractor decomposition" $\D = (H, A_0, (S_k,A_k,D_k)_{0<k<\kappa})$ of "tree-shape" $T$ with $\Sn{n}{T}=h$.
	
	We will define a "strategy" $\sigma$ for Eve in $\Reg{[1,2h]}{n+1}(G)$ that makes its choices based on the current position in $\D$ of the play of $G$ – notably, she uses its "tree-shape" and $n$-Strahler numbers of subtrees to choose registers. We then show that a "play" "consistent with" $\sigma$ is necessarily "accepting@parity accepting". More precisely, when the smallest subtree $T_\rho$ visited infinitely often has "$n$-Strahler number" $j$, we show that the maximal priority output by Eve's strategy infinitely often is $2j$.
	
	%we show that if it were to generate a dominating odd value $2j+1$ in $\Reg{[1,2h]}{n+1}(G)$, this would imply that $T_\rho$ has $n+1$ subtrees of "$n$-Strahler number" $j$, a contradiction.

	\textbf{Preliminary notations:} For $T'$ a "subtree" of $T$, we denote $D_{T'}$ the corresponding "attractor decomposition", over a subgame of vertex set $V_{T'}$. We observe that for $T'$ a "leaf" of $T$,  $D_{T'}$ is of the shape $(H',A'_0, ())$.
	We also observe that the "subtree" order $\subt$ corresponds to the inclusion order over the $D_{T'}$: we have that $D_{T_1} \subt D_{T_2}$ if and only if for $V_{T_1} \subseteq V_{T_2}$.
	Let $q$ a vertex of $G$: we denote $D(q)$ the smallest $D_{T'}$ (with $T' \subt T$) such that $q\in V_{T'}$. We denote $A(q)$ the "attractor" in $D_{T'}$ such that $q\in A(q)$ – these attractors form a partition of $V_{T'}$, hence $A(q)$ is well-defined.
	
	We recall that the usual order among "leaves" in $T$ is denoted $\prec$. That is, for $f_1,f_2$ distinct "leaves" of $T$, $f_1 \prec f_2$ if, for $T' = \langle (T'_k)_{k<\alpha}\rangle$ the smallest common ancestor of $f_1$ and $f_2$, for $T'_{k_1}$ and $T'_{k_2}$ the distinct ancestors of $f_1$ and $f_2$ respectively, $k_1 < k_2$.
	We extend this order to "attractors" in the different $(D_{T'})_{T'\subt T}$: for $A_1$ an "attractor" in $D_{T_1}$ and $A_2$ an "attractor" in $D_{T_2}$, we look at $T'$ the smallest "subtree" such that $T_1 \subt T', T_2\subt T'$. We have $D_{T'}$ of the shape $(H',A'_0, (S'_k, A'_k, D'_k)_{0<k<\kappa'})$, and thus $T'$ of the shape $\langle (T'_k)_{0<k<\kappa}\rangle$. We say that $A_1 \prec A_2$ if $\exists k_1<k_2: (A_1 = A'_{k_1} \text{ or } T_1 \subt T'_{k_1}), (A_2 = A'_{k_2} \text{ or } T_2 \subt T'_{k_2})$, or if $\exists k, T_1 \subt T'_{k}$ and $A_2 = A'_k$. We observe that this order is linear.
	Intuitively, it corresponds to the order $\prec$ on "leaves", where the attractors are seen as "leaves" of their "attractor decompositions", intertwined with the branches.
	
	We say that $q \prec q'$ if $A(q) \prec A(q')$.
	Let $q,q'$ vertices of $G$. We denote $T_{q,q'}$ the smallest common ancestor of $D(q)$ and $D(q')$. We denote $S(q,q') := \Sn{n}{T_{q,q'}}$, and $l(q,q')$ the "level" of $D_{T_{q,q'}}$.
	
	\textbf{Definition of the strategy:} Let us define the "strategy" $\sigma$ for Eve in $\Reg{[1,2h]}{n+1}(G)$, that we will then prove te be "winning". We recall that a strategy for Eve in $\Reg{[1,2h]}{n+1}(G)$ consists in \begin{itemize}
		\item a strategy in the underlying game $G$
		\item for each configuration where the move $e$ in $G$ is of odd priority $q$, the choice of an odd $i\geq p$ (else, $i:= p$)
		\item for each configuration and move in $G$, the choice of a register $r_j$.
	\end{itemize}
	The strategy $\sigma$ is defined in the following manner:
	\begin{itemize}
		\item Whenever Eve is required to play in $G$, she plays according to $\sigma_G$.
		\item After an edge $e = (q,q')$ of priority $p$ is played in $G$, if $p$ is odd and $p<l(q,q')-1$, then Eve picks $i := l(q,q')-1$, else $i := p$.
		\item After an edge $e = (q,q')$ of priority $p$ is played in $G$, if $q' \prec q$, she picks the "register" $r_0$, else if $q \prec q'$, she picks the "register" $r_{S(q,q')}$. Finally, if $A(q) = A(q')$, she picks $r_0$ if $i<l(q,q')$, else $r_1$. 
	\end{itemize} Note that these "registers" exist, as all "subtrees" of $T$ have $n$-Strahler number at most $h$. We also observe that whenever $A(q)\prec A(q')$, $p$ is even, by definition of the "attractor decomposition".
	
	Let $\rho = (\rho_l)_{l\in \NN}$ be a "play" "consistent with" $\sigma$ in $\Reg{[1,2h]}{n}(G)$. Given a transition $\rho_l$ in $\Reg{[1,2h]}{n}(G)$, for $(q,q')$ the corresponding transition in $G$, we denote $A_l := A(q')$, $p_l$ its "priority" seen in $G$, $i_l$ the "priority" chosen by Eve, $j_l$ the "register" then chosen by Eve, and $w_l$ the resulting output.
	
	We consider $T_{\rho}$, the smallest subtree of $T$ such that $(A_l)_{l\in \NN}$ eventually remains in $T_\rho$. Let $k_0$ be an index past which all the $(A_l)_{k_0<l}$ are leaves of $T_\rho$.
	
	We now prove that the sequence $(w_l)_{l\in \NN}$ is "accepting", and will prove later that Eve does not loose instantly in $\rho$.
	
	\textbf{Case $\T_\rho$ is a leaf:} If $T_\rho$ is a leaf $f$, with $D_{T_\rho} = (H_{\rho},A_{\rho}, ())$, we observe that Eve wins this play. Let $h$ be the "level" of $D_{T_{\rho}}$. As the underlying play in $G$ never sees an odd priority greater that $h$ (else it would leave $D_{T_{\rho}}$), $(i_l)_{l\in \NN}$ is eventually dominated by $h$ or a higher even priority. Furthermore, the cannot remain indefinitely in $A_{\rho}$ without seeing some edge in $H_{\rho}$ (of priority $h$) or an edge of even higher even priority. Therefore, all the $(c_{i,1})_{i \text{ odd }<h}$ always reach a value of at most 1 before being reset, as each time Eve chooses $r_1$, it is when seeing a priority greater than $i$. We also observe that infinitely often $w_l = 2$, as it is the output at each such moment. Therefore, as $r_1$ is the greatest register chosen infinitely often, and no $3$ can be output after infinitely often (as no such $c_{i,1}$ reaches $n+1$), $(w_l)_{l\in\NN}$ is winning.
	
	\textbf{Case $T_\rho$ is not a leaf:} Else, $T_\rho$ is a non-leaf node and $\rho$ alternates between different "subtrees" of $T_{\rho}$ (or between a "subtree" and its "attractor"). Let $j^\rho := \Sn{n}{T_\rho}$. Then infinitely often, the register $r_{j^{\rho}}\neq r_0$ is chosen, at each rightwards movement between "attractors" / "subtrees" of $T_{\rho}$, and no greater register is chosen past $k_0$ (as it would imply moving out of $T_\rho$, towards some other subtree such that their smallest common ancestor has "$n$-Strahler number" greater than $j^\rho$).
	
	We thus observe that infinitely often, $w_l = 2j^\rho$, as it is the default output when choosing $r_{j^\rho}$. Let us show that past $k_0$, $w_l = 2j^\rho+1$ at most $h_o$ times, for $h_o$ the number of odd priorities in $I$ below the "level" $T_\rho$'s "attractor decomposition" – and thus $(w_l)_{l\in \NN}$ is accepting. Let us suppose by contradiction that $w_l = 2j^\rho+1$ at least $h_o+1$ times after $k_0$. Then there exists some "counter" $c_{2i^\rho+1,j^\rho}$ that reaches $(n+1)+1$ twice after $k_0$.
	
	\textbf{Exhibiting $n+2$ indices at which $c_{2i^\rho +1, j^\rho}$ is incremented:} Let us look at the first time $k_1\geq k_0$ where $c_{2i^\rho +1, j^\rho}$ reaches $n+2$. At this moment, $c_{2i^\rho +1, j^\rho}=0$ and $r_{j^\rho} \neq 2^\rho+1$ (as this register was just chosen, it is updated to the current $i_l$, necessarily even). Let us look at some index $k_3> k_1$ at which $c_{2i^\rho +1, j^\rho}$ reaches $n+2$, and at $k_2$ the greatest index in $[k_1,k_3)$ such that $c_{2i^\rho +1, j^\rho}=0$ and $r_{j^\rho} \neq 2^\rho+1$ (as it is the case in $k_1$, there exists at least one such index – whenever $c_{2i^\rho +1, j^\rho}$ is reset or incremented, we can observe that after that, $r_{j^\rho}\neq 2i^\rho +1$). Therefore, $c_{2i^\rho +1, j^\rho}$ is thus incremented $n+2$ between $k_2$ and $k_3$ and is never reset. Then, $\forall l \in [k_1,k_2)$, $i_l\leq 2i^\rho+1$, and at $n+2$ different times $(l_m)_{m\in [1,n+2]} \in [k_1,k_2]^{n+2}$, the counter $c_{2i^\rho+1,j^\rho}$ is incremented (with $l_{n+2}=k_3$), that is, at such a time $l_m$, initially $r_{j^\rho}=2i^\rho+1$, and $j_{l_m} = j^\rho$.
	
	\textbf{Exhibiting $n+1$ subtrees $(T_m)_{m\in [1,n+1]}$:} We observe that for each such time $l_m$, the movement at step $l_{m}$ is increasing for the order $\prec$, and therefore $p_{l_m}$ is even (and so is $i_{l_m}$, as then $i_{l_m} = p_{l_m}$). Thus, $i_{l_m} < 2i^\rho+1$, as the latter is odd and dominates the sequence $(i_l)_{l\in [k_2,k_3)}$. Let us denote, for each such $(l_m)_{m\in [1,n+1]}$ with transition in $G$ being $(q,q')$, $T_m$ common ancestor of $A(q)$ and $A(q')$ such that $D_{T_m}$ is of "level" $2i^\rho$ – which exists: as $i_{l_m} < 2i^\rho+1$, we observe that the smallest common ancester of $A(q)$ and $A(q')$ corresponds to an "attractor decomposition" of "level" at most $2i^\rho$.
	For $m = n + 2$, it can happen that $i_{l_m}$ is even and greater that $2i^\rho + 1$, as the counters reset after the definition of the output. Therefore, this priority does not necessarily exhibit such a tree $T_{n+2}$ -- for instance, if $i_{l_m} = 2i^\rho+2$, the corresponding move might go from $T_{n+1}$ all the way to $T_1$. 
	
	Note that at each moment $l_m$, $r_{j^\rho}=2i^\rho+1$, hence there is a time $l \in (l_{m-1},l_m)$ such that $i_l = 2i^\rho+1$ (up to denoting $l_0 := k_2)$.
	Therefore, between each $l_m$, the play in $G$ leaves the current $T_m$, as $D_{T_m}$ has "level" $2i^\rho$, towards a smaller subtree of $T_\rho$ as to the order $\prec$. This thus defines $n+1$ distinct subtrees $(T_m)_{m\in [1,n+1]}$, corresponding to "attractor decompositions" of "level" at most $2i^\rho$ and "$n$-Strahler number" $j^\rho$.
	
	\textbf{The $(T_m)_{m\in [1,n+1]}$ are siblings and of $n$-Strahler number $j^\rho$:}We also observe, due to the way in which Eve chooses $i$ in the case of odd priorities, that these $(T_m)_{m\in [1,n+1]}$ are children of a same node. Indeed, each time $2i^\rho+1$ is chosen by Eve, it corresponds to $l(q,q')$ for the current $G$-transition going from $q$ to $q'$. If these $T_m$ would not be sons of a same node of "level" $2i^\rho+2$, there would be among these $2i^\rho +1$ a transition labelled by $i' > 2i^\rho+1$, thus contradicting the domination property as it would reset $c_{2i^\rho+1,j^\rho}$. 
	Therefore, the $(T_m)_{m\in[1,n+1]}$ are all "children" of a same tree $T'\subt T_\rho$, and are all of "$n$-Strahler number" $j^\rho$ (they all admit a subtree of $n$-Strahler $j^\rho$, as $\forall m \in [1,n+1], j_{l_m} = j_\rho$, and $j_\rho$ is maximal in $T_\rho$). Therefore $T'$ has "$n$-Strahler number" $j^\rho+1$. However by definition $T'$ is a subtree of $T'_\rho$ of "$n$-Strahler number" $j^\rho$, contradiction with the $n$-Strahler number being non-decreasing.
	
	\textbf{No instant-loss for Eve:} We observe that if $\Sn{n}{T_\rho} = h$, we can take $k_1 = 0$, and observe that according to the same reasoning we cannot ever have $w_l = 2h+1$, that is, Eve does not lose instantly.
\end{proof}

\end{document}